\documentclass[12pt,dvipsnames]{article}

\usepackage{iftex}
\ifLuaTeX
\usepackage{polyglossia} 
\usepackage{fontspec} 
\setmainlanguage{english}
\usepackage[nosort]{cite}
\else 
\usepackage[utf8]{inputenc}
\usepackage[T1]{fontenc}
\usepackage[nosort]{cite}
\usepackage[english]{babel} 
\fi

\usepackage[top=2cm,bottom=2cm,outer=2.5cm,inner=2.5cm,marginparwidth=2.5cm]{geometry}
\usepackage{fvextra} 
\usepackage{authblk} 

\ifLuaTeX
\usepackage[unicode]{hyperref}
\usepackage{pdftexcmds} 
\makeatletter
\let\pdfstrcmp\pdf@strcmp
\makeatother
\else
\usepackage{hyperref}
\fi

\usepackage{amsmath}
\usepackage{amsfonts}
\usepackage{amssymb}
\usepackage{empheq} 
\usepackage{romanbar} 
\usepackage{dsfont} 
\usepackage{slashed}

\usepackage{lmodern} 
\usepackage{caption}
\usepackage{subcaption} 
\usepackage{titlesec} 
\usepackage{fancyhdr} 
\usepackage{abstract} 
\usepackage[super]{nth}
\usepackage{tocloft} 

\usepackage{xcolor}

\hypersetup{
pdfstartview={FitH},
pdftitle={Anomaly of M2-branes},
pdfauthor={N. Drukker, M. Tr√©panier},
pdfsubject={},
pdfcreator={},
pdfkeywords={},
colorlinks=true,
citecolor=MidnightBlue,
linkcolor=MidnightBlue,
urlcolor=MidnightBlue
}

\numberwithin{equation}{section}
\newcommand\frontmatter{%
\clearpage
\pagenumbering{roman}
}

\newcommand\mainmatter{%
\clearpage
\pagenumbering{arabic}
}

\usepackage{mathtools}

\DeclareMathOperator{\tr}{tr}
\newcommand{\vev}[1]{\left\langle #1 \right\rangle}
\DeclareMathOperator{\vol}{vol}
\newcommand{\RicciScalar}{\mathcal{R}}
\newcommand{\IIFundForm}{\Romanbar{2}}
\newcommand{\diff}{\mathrm{d}}

\DeclareMathOperator{\Pf}{Pf}
\DeclareMathOperator*{\res}{res}

\def\bC {\mathbb{C}}
\def\bR {\mathbb{R}}
\def\cA{{\mathcal{A}}}
\def\cM{{\mathcal{M}}}
\def\cN{{\mathcal{N}}}
\def\cO{{\mathcal{O}}}
\def\cR{{\mathcal{R}}}

\newcommand{\sof}{\mathfrak{so}}

\newcommand{\bea}{\begin{eqnarray}}
\newcommand{\eea}{\end{eqnarray}}
\newcommand{\beq}{\begin{equation}}
\newcommand{\eeq}{\end{equation}}
\newcommand{\bal}{\begin{equation}\begin{aligned}{}}
\newcommand{\eal}{\end{aligned} \end{equation}}

\newcommand{\sff}{\mathrm{I\!I}}

\title{
Quantum holographic surface anomalies\\
}
\author{Nadav Drukker\thanks{\href{mailto:nadav.drukker@gmail.com}{nadav.drukker@gmail.com}}}
\author{Omar Shahpo\thanks{\href{mailto:omar.shahpo@kcl.ac.uk}{omar.shahpo@kcl.ac.uk}}}
\author[1,2]{Maxime
Tr\'epanier\thanks{\href{mailto:trepanier.maxime@gmail.com}{trepanier.maxime@gmail.com}}}
\affil[1]{\it Department of Mathematics, King's College London,\protect\\London, 
WC2R 2LS, United Kingdom}
\affil[2]{\it Department of Physics, Technion, Haifa, 32000, Israel}
\date{}

\begin{document}

\frontmatter
\maketitle

\begin{abstract}
Expectation values of surface operators suffer from logarithmic divergences 
reflecting a conformal anomaly. In a holographic setting, where 
surface operators can be computed by a minimal surface in $AdS$, the 
leading contribution to the anomaly comes from a divergence in 
the classical action (or area) of the minimal surface. We study the 
subleading correction to it due to quantum fluctuations of the minimal 
surface. In the same way that the divergence in the area does not require 
a global solution but only a near-boundary analysis, the same holds 
for the quantum corrections. We study the asymptotic form of the 
fluctuation determinant and show how to use the heat kernel to 
calculate the quantum anomaly. In the case of M2-branes describing 
surface operators in the ${\cal N}=(2,0)$ theory in 6d, our calculation 
of the one-loop determinant reproduces 
expressions for the anomaly that have been found by less direct methods.

\end{abstract}
\thispagestyle{empty}

\mainmatter
\tableofcontents

\section{Introduction}
\label{sec:intro}

Surface operators are considered rather exotic observables in quantum field 
theories, except perhaps in the context of three dimensional theories where they
appear as boundaries and interfaces, see
e.g.~\cite{AJBray_1977,ohno,Cardy:1984bb} for early work. 
The reason is probably that
they are harder to define and compute than local and line operators. And yet,
if one bothers to look for them, they are ubiquitous: they appear already in the
simplest examples of theories with scalar
fields~\cite{Metlitski:2020cqy,Rodriguez-Gomez:2022gbz,Shachar:2022fqk,Trepanier:2023tvb,
Giombi:2023dqs,Raviv-Moshe:2023yvq},
in the context of entanglement entropy in four
dimensions~\cite{Solodukhin:2008dh,Bianchi:2015liz},
in 4d supersymmetric gauge
theories~\cite{Gukov:2006jk,Gukov:2008sn,Gomis:2014eya}, in six
dimensional theories~\cite{Witten:1995zh,Strominger:1995ac,Ganor:1996nf} and
more. A simple way to engineer them is by adding extra two dimensional degrees
of freedom on a surface and coupling them in some way to 
the bulk~\cite{Gaiotto:2009fs}.

Once we define a surface operator we are faced with the task of computing its 
expectation value (or more generally correlation functions). This seems like 
a daunting problem, as it depends on the shape of the surface. In conformal 
field theories the situation is dramatically simplified, as surface operators 
are described by three \emph{anomaly coefficients}~\cite{Schwimmer:2008yh}. 
Those are similar to dimensions of local operators and central charges for 
bulk theories. Calculating 
the expectation value of a surface operator is akin to computing a two-point 
function of local operators
\beq
\label{2-point}
\vev{O(x)O(y)}=\frac{c}{|x-y|^{2\Delta_O}}\,,
\eeq
where $\Delta_O$ is the dimension of the operator. $c$ is a normalisation constant 
that in most cases is scheme dependent, with exceptions for specific operators like 
the energy momentum tensor or conserved currents. The two point function is not 
invariant under conformal transformations, but transforms in a well defined way 
determined by $\Delta_O$. In perturbative calculations of the dimension, the
quantum corrections $\Delta_O=\Delta_O^\text{cl}+\delta\Delta_O$ appear
multiplying a logarithm of a cutoff, like
\begin{align}
\log\vev{O(x)O(y)} \sim  
2\delta\Delta_O\log\epsilon\,.
\end{align}
We include the $\log$ on the left-hand side to account for the exponentiation of the
irreducible diagrams.

The anomaly coefficients of surface operators play a similar role. They appear 
in the expectation value of the surface operator multiplied by particular 
conformally covariant local densities. They govern how the observable 
transforms under conformal transformations and in explicit calculations are 
the prefactors of logarithmic divergences. For a surface operator along 
the submanifold $\bar\Sigma$ with induced metric $\bar g$, 
its expectation value is given by
\beq
\label{r-power}
\vev{O_{\bar\Sigma}}=c\,r^{-\int_{\bar\Sigma}d^2\tau\sqrt{\bar g}{\cal A} }\,,
\eeq
where $\cal A$ is known as the anomaly density.
Here $r$ is some length-scale, similar to $|x-y|$ in \eqref{2-point} 
and $c$ a scheme dependent factor. 
The bars are used to indicate quantities defined in the field theory 
or the boundary of $AdS$ and to distinguish from quantities in 
the bulk.

In practice, what we would normally find is
\beq
\log\vev{O_{\bar{\Sigma}}}\sim 
\int_{\bar\Sigma}d^2\tau\sqrt{\bar g}\,{\cal A}\,\log\epsilon\,.
\label{anomaly-integral}
\eeq
The form of ${\cal A}$ is constrained by the Wess-Zumino conditions 
to be a linear combination of independent conformally 
invariant densities on a surface~\cite{Wess:1971yu, Deser:1993yx,
Boulanger:2007st,Schwimmer:2008yh}\footnote{%
There is a second basis of conformal invariants commonly used,
where $H^2 + 4 \tr{P}$ is replaced by the traceless part of the second
fundamental form squared, $\tr\tilde{\IIFundForm}^2$. The relation between that
basis and~\eqref{eqn:anomaly} can be found in~\cite{Drukker:2020dcz}.}
\beq
\mathcal{A}_{\bar\Sigma} = \frac{1}{4 \pi} \left[
a_1 \bar\RicciScalar + a_2 (\bar H^2+4\tr \bar P) + b \tr\bar W
\right],
\label{eqn:anomaly}
\eeq
where $\bar\RicciScalar$ is the Ricci scalar of the surface, 
$\bar H^{\mu'}$ is the mean curvature of the surface, 
$\bar P$ is the pullback of the Schouten tensor 
and $\bar W$ is the pullback of the Weyl tensor to the surface. 
The geometric dependence is given by these three local densities and their 
prefactors $a_1$, $a_2$ and $b$ are the anomaly coefficients. They are 
characteristic of the operator, and should be thought of as part of the CFT 
data, just like $\Delta_O$.

Given a definition of a surface operator, we should then aim to determine 
these three numbers. One approach is to choose three geometries for which 
$\{\bar\RicciScalar\,,\,H^2 + 4 \tr{P}\,,\,\tr\bar W\}$ are 
linearly independent and determine the coefficients from the three 
examples. But given that $\cal A$ is made of local quantities, determining 
it by a local calculation should also be possible.

This was realised beautifully by Graham and Witten for the surface operators 
in the six dimensional ${\cal N}=(2,0)$ theory \cite{graham:1999pm}. 
For the $A_N$ theory at 
large $N$, the surface operators are given by M2-branes that end on 
$\bar\Sigma$ on the boundary of $AdS_7$ (or asymptotically locally $AdS_7$). 
While full classical M2-brane solutions are know in only a handful 
of examples~\cite{Maldacena:1998im,Berenstein:1998ij, mezei:2018url, 
Drukker:2021vyx, Drukker:2022beq,Drukker:2022kuz}, 
what Graham and Witten showed is that it is possible to 
solve for the near-boundary embedding and that this solution is enough in 
order to calculate the anomaly coefficients at large $N$.

The finite $N$ corrections were found only much later and by indirect methods.
They were first conjectured in~\cite{Jensen:2018rxu} based on the
holographic description of the 1/2 BPS plane in terms of bubbling
geometries~\cite{Gentle:2015jma} and the calculation of entanglement
entropy~\cite{Rodgers:2018mvq,Estes:2018tnu}.
The conjecture for $a_1$ is confirmed by $b$-extremisation~\cite{Wang:2020xkc}
and the coefficient $a_2$ can be calculated using the superconformal
index and the chiral algebra sector of the $\cN = (2,0)$
theory~\cite{Bullimore:2014upa,Chalabi:2020iie,Meneghelli:2022gps}.
The coefficient $b$ was conjectured to vanish in~\cite{Bianchi:2019sxz}
and later proven in~\cite{Drukker:2020atp} using supersymmetry.
The linear combination $a_1+2a_2$ was determined by direct calculation 
of the holographic dual of the 1/2 BPS spherical surface. First 
to leading order in $N$ using the classical 
M2-brane~\cite{Corrado:1999pi} and to first
subleading order in $N$ by evaluating 1-loop
determinants~\cite{Drukker:2020swu}.

The purpose of this paper is to rederive these results by a direct 
extension of the asymptotic analysis of Graham and Witten. 
The same philosophy still holds: 
it is enough to know the near boundary geometry of the brane to extract the 
near boundary quadratic fluctuation action and the near boundary fluctuations 
are enough to determine the anomaly coefficients.
We emphasize that the conformal anomalies obtained this way are associated with
IR divergences near the boundary of $AdS_3$---these are distinct from the
well-studied Seeley coefficients capturing UV divergences of determinants (see
e.g.~\cite{Vassilevich:2003xt}).

At the technical level we study the determinant of differential 
operators on asymptotic $AdS_3$ space. For pure $AdS_3$, this 
can be easily extracted from the 
heat kernel for scalars and spinors which is known 
exactly~\cite{Camporesi:1990wm, Giombi:2008vd,Giombi:2013fka}. For the 
purpose of deriving the anomaly, we require the first subleading 
correction of asymptotic $AdS_3$.

To do that we employ 
perturbation theory, as was used for $AdS_2$ in a somewhat different 
context in~\cite{Forini:2017whz}. The $AdS_3$ 
heat kernel times a step function serves as a Green's function for the 
heat equation, allowing for a systematic perturbative expansion. 
The first correction involves a convolution of two heat kernels with 
an extra local differential operator between them. 
Evaluating this operator and using the properties of the 
heat kernel results in expressions related to the geometric 
invariants in \eqref{eqn:anomaly} and a direct convolution 
of two heat kernels, which is easy to perform, see Section~\ref{sec:kernel}.

This formalism is valid for evaluating determinants in any context 
of asymptotic $AdS_3$ space. For the specific problem of 
the anomaly of the surface operators of the ${\cal N}=(2,0)$ theory 
we need the quadratic action for M2-branes with 
asymptotically $AdS_3$ geometry inside asymptotically $AdS_7$. 
This is derived in Appendix~\ref{app:fluct} without restriction 
to this geometry, but rather 
the quadratic fluctuation action around an arbitrary classical solution. 
This can be seen as an auxiliary result of this paper.

Applying our heat kernel technology to these differential operators, 
we evaluate their determinants for both bosonic and fermionic 
modes and extract the anomaly coefficients
\beq
\label{coefficients}
a_1=\frac{1}{2}+{\cal O}(1/N)\,,\qquad
a_2=-N+\frac{1}{2}+{\cal O}(1/N)\,,\qquad
b={\cal O}(1/N)\,.
\eeq
These expressions agree with the existing literature~\cite{Jensen:2018rxu, 
Rodgers:2018mvq,Estes:2018tnu, Wang:2020xkc, 
Bullimore:2014upa,Chalabi:2020iie,Meneghelli:2022gps, 
Bianchi:2019sxz, Drukker:2020atp, Drukker:2020swu}.

\section{Asymptotic \texorpdfstring{$AdS_3\subset AdS_7$}{$AdS3 < AdS7$}}
\label{sec:3metrics}

In this section we set up the asymptotic $AdS$ geometries that play a role in the 
calculation. We first review the work 
of Graham and Witten~\cite{graham:1999pm} finding a near-boundary classical 
brane solution. We then implement a change of coordinates to bring also 
the induced metric on the brane to Fefferman-Graham 
form~\cite{fefferman1985conformal}, which simplifies 
the calculation of the determinants in the next section.

\subsection{Asymptotic $AdS_7$ geometry}
\label{sec:GW}

Following~\cite{graham:1999pm,Drukker:2020dcz}, we look at a bulk geometry asymptotic to 
$AdS_7\times S^4$. We choose $y$ as the coordinate normal to the boundary, 
such that the asymptotic form of the metric is~\cite{graham:1999pm,Drukker:2020dcz}
\beq
\label{metric-FG}
G=\frac{R^2}{y^2}\left(\diff y^2+\bar G+y^2 \bar{G}^{(1)}\right)
+\frac{R^2}{4}G_{S^4}+\cO(y^2)\,.
\eeq
$\bar G$ is the metric on the boundary of space and $\bar G^{(1)}$ is
fixed by the (super)gravity equations to be the Schouten tensor of $\bar G$
\beq
\label{G1}
\bar G_{MN}^{(1)}=-\bar P_{MN}(\bar G)\,.
\eeq
Here we use $MN$ for coordinates on all of (asymptotically) $AdS_7$.

We study the embedding of a 3d M2-brane with world-volume 
$\Sigma$ into this geometry, where the brane ends
along a 2d surface $\bar\Sigma$ in the 6d boundary of
asymptotically locally $AdS_7$. We use coordinates $\tau^a$ with $a=1,2$ on the
2d surface and $\sigma^\mu$ with $\mu=1,2,3$ and $\sigma^3=y$ on the M2-brane
world-volume. The asymptotically locally $AdS_7$ space is parametrised by
coordinates $x^M$. We take three to be $x^\mu = \sigma^\mu$, and we
require that the remaining coordinates $x^{\mu'}(\sigma)$ are orthogonal to
$\bar{\Sigma}$ at the boundary.
Finally we parametrise the $S^4$ by $z^i$ with 
$i =1,\cdots,4$ and we restrict to classical solutions localised 
at a point on $S^4$
$z^i(\sigma)=0$, representing the north pole of $S^4$.

The bosonic part of the M2-brane 
action~\cite{Bergshoeff:1987cm} is the volume form of the
induced metric and the pullback of the three-form $A_3$
\begin{align}
\label{action}
 S_\text{M2} =
 T_\text{M2}
 \int_\Sigma \left( \vol_\Sigma + i A_3 \right)\,.
\end{align}
Here $A_3$ is the potential for the flux
\begin{align}
 F_4=dA_3 = \frac{3}{8} R^3 \vol_{S^4} +\,\cO(y^2)\,,
 \label{eqn:f4}
\end{align}
where $\vol_{S^4}$ is the volume of the unit sphere, as in 
\eqref{metric-FG}. As mentioned above, we assume the surface is localised 
at a point in $S^4$, so the pullback of $A_3$ in \eqref{action} vanishes.

The equations of motion for the brane are those of a minimal surface, 
which can be elegently expressed as the vanishing of the mean curvature 
vector $H^{M}$. Recalling some definitions, the second fundamental form is
\beq
\label{SFF1}
\sff^{M}_{\mu\nu}
=(\partial_\mu\partial_\nu x^{P}
+\partial_\mu x^{Q}\partial_\nu x^{R}\Gamma^{P}{}_{QR})
(\delta^{M}_{P}-\partial^\rho x^{M}\partial_\rho x^{N}G_{NP})\,,
\eeq
and using the inverse of the induced metric $g_{\mu\nu}$ we get 
the mean curvature vector as
\beq
H^{M}=g^{\mu\nu}\sff^{M}_{\mu\nu}\,.
\eeq

In fact, if the coordinates $x^{\mu'}$ are orthogonal to the brane not just at
the boundary but everywhere, 
then the second fundamental form simplifies and the only nonzero components are
\beq
\label{SFF2}
\sff^{\mu'}_{\mu \nu}= \left.-\frac{1}{2}G^{\mu'\nu'}\partial_{\nu'}G_{\mu \nu}
\right|_\Sigma\,.
\eeq

We can express $H^{\mu'}$ in a Fefferman-Graham expansion as 
power series in $y$ in \eqref{metric-FG}. If $x^{\mu'}$ were constant, 
then it would be the same as the mean curvature on the boundary surface 
$\bar\Sigma$, so $H^{\mu'}=\bar H^{\mu'}$ (or strictly speaking, the
pullback of $\bar H^{\mu'}$, since these two objects are in 
different bundles).
If $x^{\mu'}$ is not a constant, we find 
to lowest nontrivial order in $y$~\cite{Drukker:2020dcz}
\beq
\label{x(y)}
H^{\mu'}=\bar H^{\mu'}+y^3\partial_y(y^{-3}\partial_yx^{\mu'})
+\cO(y^2)\,.
\eeq
Imposing the equations of motion sets this to zero and fixes
$x^{\mu'}=\bar H^{\mu'} y^2/4$, as found in \cite{graham:1999pm}.

Before imposing the equations of motion and keeping only terms of order
$\cO(y^0)$, the induced metric on the world-volume is
\bal
g_{yy}&=\frac{R^2}{y^2}\left(1
+\partial_y x^{\mu'}\partial_y x^{\nu'}
\bar{g}_{\mu'\nu'}
\right).
\\
g_{ab}&=\frac{R^2}{y^2}\left(\bar g_{ab}
-y^2\bar P_{ab}
-2\bar\sff^{\mu'}_{ab}x^{\nu'} \bar{g}_{\mu'\nu'}
\right),\\
g_{ay}&=0\,,
\eal
Here $\bar g_{ab}$ is the metric on $\bar\Sigma$ and $\bar P_{ab}$ is the
pullback of the bulk Schounten tensor to the brane, and likewise for the second
fundamental form. We also use $g_{\mu'\nu'} = G_{\mu'\nu'}|_{\Sigma}$ for the metric
evaluated on the brane, and $\bar{g}_{\mu'\nu'} = G_{\mu'\nu'}|_{\bar{\Sigma}}$
for its value at the boundary of $AdS$.

Plugging in the solution to the asymptotic equations, 
$x^{\mu'}=\bar H^{\mu'} y^2/4$, gives the induced metric
\bal
\label{asym-met-1}
g_{yy}&=\frac{R^2}{y^2}\left(1+\frac{y^2}{4}\bar H^2\right),
\\
g_{ab}&=\frac{R^2}{y^2}\left(\bar g_{ab}
-\bar P_{ab}y^2
-\frac{y^2}{2}\bar\sff^{\mu'}_{ab}\bar H^{\nu'}
\bar{g}_{\mu'\nu'}\right),\\
g_{ay}&=0\,.
\eal

The classical action (expanding the integrand to order $y^{-1}$) is now
\bal
\label{S-classical}
S_\text{classical}
&= T_\text{M2}\int_\Sigma\diff^3\sigma\,\frac{R^3}{y^3}
\sqrt{\bar g\left(1+\frac{y^2}{4}\bar H^2\right)
\left(1-y^2\tr\bar P-\frac{y^2}{2}\bar H^2\right)}
\\
&= T_\text{M2}R^3\int_0^\infty \frac{\diff y}{y^3}
\int_{\bar\Sigma}\diff^2\tau\,\sqrt{\bar g}
\left(1-\frac{y^2}{8}\bar H^2-\frac{y^2}{2}\tr\bar P\right)\,.
\eal
Using $T_\text{M2}R^3=2N/\pi$, we get a quadratic and logarithmic divergences
\beq
S_\text{classical}
=\frac{N}{\pi\epsilon^2}\vol(\bar\Sigma)
+\frac{N}{4\pi}\int_{\bar\Sigma}\diff^2\tau
\sqrt{\bar g}\left(\bar H^2+4\tr\bar P\right)\log\epsilon\,.
\label{eqn:Sclassical}
\eeq
The quadratic divergence can be cancelled by an appropriate Legendre 
transform~\cite{Drukker:1999zq, Mori:2014tca}. Evaluating 
$\exp[-S_\text{classical}]$ then gives a power of $1/\epsilon$ 
which should be minus the anomaly \eqref{r-power}, so we 
can identify the leading result 
at large $N$ for the anomaly coefficients, namely $a_1=b=0$ and 
$a_2=-N$ as in \eqref{coefficients} \cite{graham:1999pm}.

\subsection{Asymptotic $AdS_3$ brane geometry}
\label{sec:AdS3}

The induced metric~\eqref{asym-met-1} is perfectly fine in order to 
plug into the action and evaluate the classical anomalies. The coordinates 
and metric have 
some issues that make them less than ideal for the quantum calculation 
in Section~\ref{sec:det}. First, the induced metric is not in 
Fefferman-Graham form, so the asymptotic 
$AdS_3$ structure needed in there is not manifest.
Second, the quadratic fluctuation action derived in Appendix~\ref{app:fluct} 
assumes coordinates tangent and normal to the brane. The fact that 
$x^{\mu'}$ depends on $y$ means that they are not orthogonal. 

Both of those issues can be resolved with a change of coordinates 
to $z$ and $u^{\mu'}$ defined via
\bal
 x^{\mu'} &= u^{\mu'} + \frac{y^2}{4} \bar{H}^{\mu'}\,,
\\
 y &= z \exp{\left( -\frac{1}{2} x^{\mu'} \bar{H}_{\mu'} 
 - \frac{z^2}{16} \bar{H}^2 \right)}\,.
\label{zu}
\eal
The coordinates $\tau^a$ remain untouched for now. 
The choice of $u^{\mu'}$ is such that it vanishes on the classical solution 
\eqref{x(y)} 
and then the definition of $z$ makes the metric block diagonal and simplifies 
the induced metric. Plugging this into \eqref{metric-FG}, we find to order 
$z^0$
\bal
\label{new-FG}
 \frac{G}{R^2}
 &=
 \frac{\diff z^2}{z^2}
 + \frac{e^{u^{\rho'} \bar{H}_{\rho'}}}{z^2} \left[ \left( \bar{G}_{ab} \diff \tau^a
 \diff \tau^b + \bar{G}_{\mu'\nu'} \diff u^{\mu'} \diff u^{\nu'} \right)
 \left( 1 + \frac{z^2}{8} \bar{H}^2 \right)
 + 2 \bar{G}_{a \mu'} \diff \tau^a \diff u^{\mu'}
 \right]\\
 &\quad{}- \frac{1}{2} \bar{H}_{\mu'} \bar{H}_{\nu'} \diff u^{\mu'} \diff u^{\nu'}
 + \bar{G}_{a \mu'} \left( \frac{\diff z \diff \tau^{a}}{z} \bar{H}^{\mu'} -
 \frac{\diff \tau^a \diff u^{\nu'}}{2} \bar{H}^{\mu'} \bar{H}_{\nu'} \right)
 -\bar P + \frac{G_{S^4}}{4}\,.
\eal
Like \eqref{metric-FG}, this metric is also in Fefferman-Graham form. 
Indeed, the 
Fefferman-Graham expansion is unique only given a boundary metric $\bar G$. 
The terms in the square bracket in the first line of~\eqref{new-FG} 
(excluding $z^2\bar H^2/8$) are exactly $\bar G$, so 
we see that we conformally transformed 
the boundary metric by $\exp[u^{\rho'}\bar H_{\rho'}]$. Thus~\eqref{new-FG} 
is the Fefferman-Graham expansion for a 
different choice of boundary metric in the same conformal class as $\bar G$. 

All the terms of order $z^0$ (with the exception of the $S^4$ part) are 
minus the Schouten tensor for the conformally transformed boundary metric. 
Of course, given that now $u^{\mu'}=0$ on the classical solution, the mean 
curvature for the surface $\bar\Sigma$ in this conformally transformed 
metric vanishes.

We can find the induced metric by setting $u^{\mu'}=0$, where also 
$\bar g_{a\mu'}=0$, resulting in
\beq
\label{induced-FG}
 \frac{g}{R^2}
 =
 \frac{\diff z^2}{z^2}
 + \frac{\diff \tau^a \diff \tau^b}{z^2} \left(
 \bar{g}_{ab} - z^2 \bar\Pi_{ab} 
 \right).
\eeq
where we defined
\beq
\label{Pi}
\bar\Pi_{ab}= 
 - \frac{1}{8} \bar{H}^2 \bar{g}_{ab} 
 + \frac{1}{2} \bar{\IIFundForm}^{\nu'}_{ab} \bar{H}^{\mu'} \bar{g}_{\mu'\nu'}
 +\bar{P}_{ab}\,,
\qquad
\tr\bar\Pi=\frac{1}{4}H^2+\tr\bar P\,.
\eeq

The induced metric~\eqref{induced-FG} is now also in Fefferman-Graham form 
for an asymptotically locally $AdS_3$ with boundary metric $\bar g$ 
on $\bar\Sigma$. Note that the correction $z^2\bar\Pi$ is not the 
Schouten tensor of $\bar g$, as in the bulk case \eqref{G1}, 
since we do not impose 
an Einstein equation on the world-volume, but rather the minimal 
surface equations.

In fact, we could have started the analysis from this statement
\begin{quote}\it
 Given a surface $\bar\Sigma\subset\bar M_6$ and a conformal 
 class on $\bar M_6$, 
 one can choose a metric in that class such that $\bar\Sigma$ is minimal,
 so $\bar{H}^{\mu'}=0$. The Graham-Witten solution then is $u^{\mu'}=0$ 
 and the induced metric~\eqref{asym-met-1} is automatically in 
 Fefferman-Graham form.%
\footnote{The discussion above makes this statement manifest and 
despite asking several experts, we could not find a reference 
that discusses this in any detail.} 
\end{quote}
Note that the conformal transformation used in \eqref{zu}, \eqref{new-FG} is 
defined only locally, but that shouldn't matter for our calculation of 
a local anomaly density.

In practical term, this other approach would set $\bar H^{\mu'}=0$ in 
all calculations, so $\bar\Pi_{ab}=\bar P_{ab}$. 
We could also work in this setting and 
replace $\tr\bar P\to \bar H^2/4+\tr\bar P$ in the final expressions, 
as this is the combination that appears in the anomaly~\eqref{eqn:anomaly}. 
We did perform the calculation keeping nonzero $\bar H^{\mu'}$ just to 
make sure that we do not make any mistakes.

\section{Determinants on asymptotically $AdS_3$}
\label{sec:det}

Having reviewed the classical M2-brane solution and presented a 
convenient form for the metric with explicit asymptotically $AdS_3$ induced 
metric on the brane, we now develop the tools to evaluate determinants 
of differential operators on such submanifolds. In turn, we apply this 
to the semiclassical M2-brane, where the action for quadratic 
fluctuations is evaluated in Appendix~\ref{app:fluct}.

In the absence of a full classical solution one cannot expect to be able 
to fully evaluate the determinant, but we are interested only in the 
logarithmically divergent terms that contribute to the anomaly and arise from IR
divergences near the boundary of $AdS$. 
We rely on the heat kernel method to evaluate the determinants and 
apply them only in the near boundary region to first nontrivial order 
beyond pure $AdS_3$.

The determinant of a differential operator $L$ can be, in principle, calculated
from the heat kernel $K(t;\sigma,\sigma_0)$ satisfying
\begin{align}
 (L_{\sigma} + \partial_t) K(t;\sigma,\sigma_0) = 0\,, \qquad
 \lim_{t \to 0} K(t;\sigma,\sigma_0) = \frac{1}{\sqrt{g}}
 \delta^{(3)}(\sigma -\sigma_0)\,.
 \label{eqn:heatkerneleqn}
\end{align}
The subscript $\sigma$ here is meant to emphasize that the differential 
operator acts on the point $\sigma$.
If one can solve the heat kernel equation, the determinant of the operator
is then obtained as
\begin{align}
 \log\det(L) = - \int_\Sigma \vol_\Sigma \int_0^\infty \frac{\diff t}{t}
 \lim_{\sigma \to \sigma_0}
 K(t;\sigma,\sigma_0).
 \label{eqn:detheatkernel}
\end{align}
For more details on the heat kernel and its applications, see the
review~\cite{Vassilevich:2003xt}.

We proceed now to study heat kernels on asymptotically $AdS_3$ for 
the massive scalar laplacian, for vector bundles and for spinors. In the case 
of the M2-brane that we are interested in, there are four massless scalars from 
the fluctuations in $S^4$, four scalars for fluctuations in $AdS_7$, which see
the nontrivial geometry of the normal bundle of the M2-brane so should be treated as taking
value in an $SO(4)$-vector bundle, and finally there are 16 fermi fields. See
Appendix~\ref{app:fluct} for the derivation of the quadratic action.

\subsection{Heat kernel asymptotics}
\label{sec:kernel}

In practice, solving the heat kernel equation for an arbitrary kinetic 
operator $L$ on an arbitrary manifold is impossible. 
Fortunately, for the purpose of extracting anomaly 
coefficients, we only need the
behavior of the differential operators near the boundary of $AdS$.

We start with a massive scalar laplacian $L = -\Delta + M^2$ 
on the asymptotically locally $AdS_3$ world-volume 
given in \eqref{induced-FG} and further simplify the calculation 
by choosing Riemann normal coordinates for the metric 
$\bar g$ about a particular point 
(corresponding to $\tau_0=(0,0)$ and at fixed $z_0$), such that
\bal
\label{asym-met-3}
g_{zz}
&=\frac{R^2}{z^2}\,,
\\
g_{az} &= 0 \,,
\\
g_{ab}
&=\frac{R^2}{z^2}\delta_{ab}
-R^2\left(\frac{1}{3z^2}\bar \cR_{acbd}\tau^c\tau^d 
+\bar \Pi_{ab} 
\right).
\eal
For simplicity we set the $AdS$ radius $R = 1$ in the following.
We proceed by considering the parenthesis on the last line as a perturbation 
about the $AdS_3$ geometry. More precisely, 
we treat $(\tau^a,z)$ as homogeneous coordinates 
and expand in a power series in them, so
$g_{\mu\nu}=g_{\mu\nu}^{(-2)}+g_{\mu\nu}^{(0)}+\cdots$ 
with the index indicating the degree and
\bal
\label{expanded-metric}
g_{\mu\nu}^{(-2)}\diff\sigma^\mu\diff\sigma^\nu
&=\frac{1}{z^2}(\diff z^2+\delta_{ab}\diff \tau^a \diff\tau^b)\,,
\\
g_{\mu\nu}^{(0)}\diff\sigma^\mu\diff\sigma^\nu
&=-\left(\frac{1}{3}\bar \cR_{acbd} \frac{\tau^c\tau^d}{z^2}
+\bar \Pi_{ab} 
\right)\diff\tau^a\diff\tau^b\,.
\eal
Likewise, the determinant of the metric is expanded as
\bal
\sqrt{g}^{(-3)}&=\frac{1}{z^3}\,,
\\
\sqrt{g}^{(-1)}
&=-\frac{1}{z}\left(\frac{1}{6}\bar{\RicciScalar}_{cd}\frac{\tau^c\tau^d}{z^2}
+\frac{1}{2} \tr\bar\Pi\right).
\label{eqn:expanded-det-metric}
\eal

We now expand the kinetic operator $L=-\Delta+M^2$ in the same way as
\beq
L=L^{(0)}+L^{(2)}+\cdots
\eeq
For a massive field, we assume an expansion of the mass term 
$M^2 = \cM^{(0)}
+ \cM^{(2)} + \cdots$, and the kinetic operator takes the form
\bal
L^{(0)}
&=-\frac{1}{\sqrt{g}^{(-3)}}
\partial_\mu \sqrt{g}^{(-3)} g^{(2)\,\mu\nu}\partial_\nu + \cM^{(0)}\,,
\\
L^{(2)}
&=
\frac{\sqrt{g}^{(-1)}}{\left(\sqrt{g}^{(-3)} \right)^2}
\partial_\mu \sqrt{g}^{(-3)} g^{(2)\,\mu\nu}\partial_\nu
-\frac{1}{\sqrt{g}^{(-3)}}
\partial_\mu\left( \sqrt{g}^{(-1)} g^{(2)\,\mu\nu}
+\sqrt{g}^{(-3)} g^{(4)\,\mu\nu}\right)\partial_\nu
+ \cM^{(2)}\,,
\label{eqn:L2abstract}
\eal
where of course $g^{(4)\,\mu\nu}= -g^{(2)\,\mu \rho} g_{\rho \sigma}^{(0)}
g^{(2)\,\sigma \nu}$.

Plugging in the metric \eqref{expanded-metric} yields
\bal
L^{(0)}
&=-z^3 \partial_{z} \left( \frac{1}{z} \partial_{z}\right)
- z^2\delta^{ab} \partial_a \partial_b + \cM^{(0)} \,,
\\
L^{(2)}
&=\tr\bar{\Pi}\, z^3 \partial_z
 + \frac{2 z^2}{3} \bar{\RicciScalar}_{ab} \tau^a \partial_b
 -\left( \frac{z^2}{3} \bar{\RicciScalar}_{acbd} \tau^c \tau^d + z^4 \bar\Pi_{ab} \right)
 \partial_a \partial_b
+ \cM^{(2)}\,.
\label{eqn:explicitdiffopexp}
\eal
We emphasize that the kinetic operator $L$ acts on the point
$\sigma$, and is evaluated at that point. The quantities 
($\cM, \bar\Pi, \cdots$) appearing in $L^{(2)}$ are all evaluated at $\sigma_0$.

The heat kernel itself can also be expanded as
\beq
K(t;\sigma,\sigma_0)
=K^{(0)}(t;\sigma,\sigma_0)
+K^{(2)}(t;\sigma,\sigma_0)
+ \cdots
\label{eqn:expheatkernel}
\eeq
To calculate the anomaly we only need those first two terms. $K^{(0)}$ is of 
homogeneous degree zero, so the volume integral in \eqref{eqn:detheatkernel} 
has a quadratic divergence. It may also contribute logarithmic divergences 
from the subleading terms in the volume form 
$\sqrt{g}^{(-1)}$ \eqref{eqn:expanded-det-metric}. 
$K^{(2)}$ is quadratic, so like the 
$y^2$ terms in the classical action \eqref{S-classical}, gives 
logarithmic divergences contributing to the conformal anomaly.

The heat kernel equation~\eqref{eqn:heatkerneleqn} then reduces to the 
recursive relations
\begin{align}
 (L^{(0)} + \partial_t) K^{(0)} = 0\,,
 \qquad
 (L^{(0)} + \partial_t) K^{(2)} = -L^{(2)} K^{(0)}\,.
 \label{eqn:heatkerneleqnexp}
\end{align}
The first equation is the $AdS_3$ heat kernel equation, its solution is given
by~\cite{Camporesi:1990wm} (also~\cite{Giombi:2008vd,Giombi:2013fka})
\begin{align}
\label{K0}
K^{(0)}(t;\sigma,\sigma_0) =
\frac{1}{(4\pi t)^{3/2}}
\frac{\rho}{\sinh \rho}
\exp\left( -\left(1+\cM^{(0)}\right)t - \frac{\rho^2}{4t} \right) \,,
\end{align}
where $\rho$ is the geodesic distance in $AdS_3$ between $\sigma$ and $\sigma_0$
\begin{align}
 \rho =
 \mathrm{arccosh}\left( \frac{z^2 + z_0^2 + |\tau - \tau_0|^2}{2 z z_0}
 \right)\,.
 \label{eqn:geodesicdist}
\end{align}

The equation for $K^{(2)}$ can be solved by observing that 
$K^{(0)}(t' - t;\sigma,\sigma_0)\Theta(t'- t)$ 
(with $\Theta$ the Heaviside step function) is a Green's function for 
$L^{(0)}+\partial_t$, i.e.
\begin{align}
(L^{(0)} + \partial_t) K^{(0)}(t'-t;\sigma,\sigma_0) \Theta(t'-t)
= \frac{1}{\sqrt{g}^{(-3)}}\delta^{(3)}(\sigma-\sigma_0)\delta(t' - t)\,.
\end{align}
This allows us to express $K^{(2)}$ as the integral
\bal
 K^{(2)}(t;\sigma,\sigma_0)
 &=
 -\int_0^t \diff t' \int_\Sigma \diff^3 \sigma'\sqrt{g}^{(-3)}
 K^{(0)}(t - t';\sigma,\sigma') L^{(2)}_{\sigma'}
 K^{(0)}(t';\sigma',\sigma_0)\\
&\quad 
-\frac{\sqrt{g}^{(-1)}}{\sqrt{g}^{(-3)}}K^{(0)}(t;\sigma,\sigma_0) \,,
 \label{eqn:K2int}
\eal
where the second line ensures that 
\beq
\lim_{t \to 0} K^{(2)}(t;\sigma,\sigma_0) 
=-\frac{\sqrt{g}^{(-1)}}{\big(\sqrt{g}^{(-3)}\big)^2}\delta^{(3)}(\sigma-\sigma_0)\,,
\eeq
so that the boundary conditions at $t = 0$~\eqref{eqn:heatkerneleqn} are 
compatible with the subleading terms in the measure $\sqrt{g}$ 
\eqref{eqn:expanded-det-metric}.

To evaluate the determinant we need \eqref{eqn:detheatkernel} 
the coincident point limit of the heat kernel $K(t;\sigma_0,\sigma_0)$, 
also known as the trace of the heat kernel. For $K^{(0)}$, we 
have the explicit expression \eqref{K0} from which we get
\beq
\label{coincident}
\lim_{\sigma \to \sigma_0}
K^{(0)}(t;\sigma,\sigma_0) =\frac{1}{(4\pi t)^{3/2}}
\exp\left( -\left(1+\cM^{(0)}\right)t\right)\,.
\eeq

For $K^{(2)}$ we can use the integral expression~\eqref{eqn:K2int} to evaluate
the coincident limit. $K^{(0)}$ depends on the coordinates only through the geodesic distance
$\rho(\sigma',\sigma_0)$, so it is natural to change coordinates for $AdS_3$ to
\begin{align}
 \diff s^2 = \left(\diff \rho^2 + 
 \sinh^2{\rho}(\diff\theta^2+\sin^2\theta\,\diff\phi^2)\right)\,.
 \label{eqn:AdS3}
\end{align}
The change of variables is given by
\begin{align}
 z =
 \frac{z_0}{\cosh{\rho} - \sinh \rho \cos \theta}\,,
 \qquad
 \tau_a =
 \frac{z_0 \sinh \rho \sin{\theta}\, e_a}{\cosh{\rho} - \sinh{\rho} \cos{\theta}}\,,
 \label{eqn:ads3changeofvariables}
\end{align}
where $e_a=(\cos\phi,\sin\phi)_a$.

We now can express $L^{(2)}$ \eqref{eqn:explicitdiffopexp} 
in these coordinates, but since we act with it on 
$K^{(0)}$, we only need to keep derivatives with respect to $\rho$. Those are
\begin{equation}
\begin{aligned}
 L^{(2)}f(\rho)
 =
 -z_0^2 &\left[
 \frac{\bar{\RicciScalar} \sinh{\rho} \sin^2{\theta}}
 {6 (\cosh{\rho} - \sinh{\rho} \cos{\theta})^3} \partial_\rho
 +
 \tr\bar\Pi
 \frac{\coth{\rho}}{\left( \cosh{\rho} - \sinh{\rho} \cos{\theta} \right)^2}
 \partial_\rho \right.\\
 &\ \ + \left. 
 \bar{\Pi}_{ab} 
 \frac{\sin^2{\theta}\, e_a e_b}{\left(\cosh{\rho} - \sinh{\rho} \cos\theta\right)^4}
 \left( \partial_\rho^2 - \coth{\rho} \partial_\rho \right) \right]f(\rho)
 +\cM^{(2)}f(\rho)
\,.
\end{aligned}
\end{equation}

The integral in~\eqref{eqn:K2int} is
\begin{align}
 - \int_0^t \diff t' \int_\Sigma \diff \rho \,\diff \theta \,\diff \phi \sinh^2{\rho}
 \sin{\theta}
 K^{(0)}(t-t';\rho) L^{(2)} K^{(0)}(t';\rho)\,.
\end{align}
The $\phi$ integral is easy to do using
\begin{align}
\label{ang-int}
 \int_0^{2\pi} \diff \phi\, e_a e_b = \pi \delta_{ab}\,.
\end{align}
The $\theta$ integral can be done as well to get
\begin{equation}
\begin{aligned}
 -4\pi
 & \int_0^t \diff t' \int_0^\infty \diff \rho \sinh^2{\rho}\,
 K^{(0)}(t-t';\rho) \\
 & \times\left[ z_0^2 \left(-\frac{1}{3}\tr\bar\Pi
 \left( \partial_\rho^2 + 2 \coth{\rho} \partial_\rho \right) +
 \frac{\bar{\RicciScalar}}{6} \frac{\cosh{\rho} \sinh{\rho} - \rho}{\sinh^2{\rho}}
 \partial_\rho \right)
+ \vev{\cM^{(2)}}
\right]
 K^{(0)}(t';\rho)\,.
\end{aligned}
\label{eqn:integratedL2scalar}
\end{equation}
$\vev{\cM^{(2)}}$ denotes the average of $\cM^{(2)}$ over the spherical
coordinates.
We can simplify further the coefficient of $\tr\bar\Pi$ by using the heat kernel
equation~\eqref{eqn:heatkerneleqnexp}, which in these coordinates is
\begin{align}
 \left( -\partial_\rho^2 - 2 \coth{\rho}\, \partial_\rho + \cM^{(0)} + \partial_t' \right)
 K^{(0)}(t';\rho) = 0\,,
\end{align}
which removes both first and second order $\rho$ derivatives 
and we obtain
\begin{equation}
\begin{aligned}
\label{D-integral}
 4\pi &
 \int_0^t \diff t' \int_0^\infty \diff \rho \sinh^2{\rho}\,
 K^{(0)}(t-t';\rho)\\
 &\times \left[
 \frac{z_0^2}{3} \tr\bar\Pi\, \left(\partial_{t'} + \cM^{(0)} \right)
 - \frac{z_0^2}{6}\bar{\RicciScalar} \frac{\cosh{\rho} \sinh{\rho} - \rho}{\sinh^2{\rho}} \partial_\rho
 - \vev{\cM^{(2)}}
 \right] K^{(0)}(t';\rho)\,.
\end{aligned}
\end{equation}

The remaining integrals can be evaluated without much difficulty 
after plugging in the expression for $K^{(0)}$ in \eqref{K0}. 
A more elegant way to evaluate them is to use partial integration to reduce them
to a convolution of heat kernels. Consider first the term proportional to
$\tr{\bar{\Pi}}$. We can take out the $t'$ derivative as
\bal
&
\frac{4\pi z_0^2}{3} \tr\bar\Pi\,
\int_0^t \diff t' \int_0^\infty \diff \rho \sinh^2{\rho}
K^{(0)}(t - t';\rho) 
\left(\partial_{t'} + \cM^{(0)} \right) 
K^{(0)}(t';\rho)
\\
&=
\frac{4 \pi z_0^2}{3}\tr\bar\Pi \int_0^t \diff t' 
\left(\partial_{t'}+\partial_t +
\cM^{(0)} \right)
\int_0^\infty \diff \rho \sinh^2{\rho}
K^{(0)}(t - t';\rho) 
K^{(0)}(t';\rho)\,.
\eal
Now the $\rho$ integral is the convolution of two heat kernels (up to the
factor $4\pi$ from the spherical integral), 
giving simply $K^{(0)}(t;\sigma_0,\sigma_0)$.
The $t'$ dependence completely drops out and the integral over 
it gives $t$, leaving
\begin{align}
 \frac{z_0^2}{3} \tr\bar\Pi\, 
 t (\partial_t + \cM^{(0)}) K^{(0)}(t;\sigma_0,\sigma_0)\,.
 \label{eqn:k2trPiint}
\end{align}
Second, for the term proportional to $\bar\RicciScalar$, with 
an exchange of $t'\to t-t'$ the derivative can be moved to the 
other $K^{(0)}$, so we can write it as
\bal
&-\frac{\pi z_0^2}{3} \bar{\RicciScalar}
 \int_0^{t} \diff t' \int_0^\infty \diff \rho
 \left( \cosh{\rho} \sinh{\rho} - \rho \right) 
 \partial_\rho\left(K^{(0)}(t - t';\rho) K^{(0)}(t';\rho)\right)
 \\&=
 \frac{2\pi z_0^2}{3} \bar{\RicciScalar}
 \int_0^{t} \diff t' \int_0^\infty \diff \rho \sinh^2{\rho}\,
 K^{(0)}(t - t';\rho)
 K^{(0)}(t';\rho)
 =\frac{z_0^2}{6} \bar{\RicciScalar} t K^{(0)}(t;\sigma_0,\sigma_0)\,.
 \label{eqn:k2Ricciint}
\eal
The last term in \eqref{D-integral} is that proportional to 
$\cM^{(2)}$, and assuming $\vev{\cM^{(2)}}$ doesn't depend on $\rho$ (as we show
in the case of interest below), it is directly the convolution of the two heat kernels.

Adding~\eqref{eqn:k2trPiint}, \eqref{eqn:k2Ricciint} and the contribution
from $\cM^{(2)}$ and plugging the value of the heat kernel at coincident points
\eqref{coincident} 
gives
\begin{align}
 \left[ -\frac{z_0^2}{3} \tr\bar\Pi\, 
 \left(\frac{3}{2t}+1\right)
 + \frac{z_0^2}{6} \bar{\RicciScalar}
 + \vev{\cM^{(2)}}
 \right]
 \frac{\exp\left( -(1+\cM^{(0)})t \right)}
 {(4 \pi)^{3/2} t^{1/2}}\,.
\end{align}
Finally, adding the term which corrects the boundary conditions from 
the second line of \eqref{eqn:K2int}, it exactly cancels the 
$3/2t$ term and we get
\begin{align}
 K^{(2)}(t;\sigma_0,\sigma_0)
 =
 \left[
 \frac{z_0^2}{6}(\bar{\RicciScalar} - 2 \tr\bar\Pi)
 - \vev{\cM^{(2)}}
 \right]
 \frac{\exp\left( -(1+\cM^{(0)})t \right)}
 {(4 \pi)^{3/2} t^{1/2}}\,.
 \label{eqn:k2scalarres}
\end{align}

Note the appearance of the combination $\bar{\RicciScalar} - 2 \tr{\bar{\Pi}}$.
The correction to the $AdS_3$ heat kernel coming from geometric terms is
expected to be proportional to this combination (up to perhaps also a contribution
from terms involving the Weyl tensor), because of the following 
consistency check.
Take the surface to be a sphere $\bar{\Sigma} = S^2$,
with corresponding classical geometry $AdS_3$~\cite{Berenstein:1998ij}. In that
case the heat kernel is exactly $K^{(0)}$, so we should find $K^{(2)} = 0$.
Indeed, although both $\bar{\RicciScalar} = 2$ and $\tr{\bar{\Pi}} = 1$ are
nonzero for the sphere, the combination $\bar{\RicciScalar} - 2 \tr{\bar{\Pi}}$
vanishes.

For the evaluation of the determinant~\eqref{eqn:detheatkernel}, 
we need the $t$ integral of the trace of the heat kernel, 
\eqref{coincident} and \eqref{eqn:k2scalarres},
which are simple gamma function integrals. Reinstating powers of $R$ we get
\begin{align}
 \label{t-int-0}
 \int_0^\infty \frac{\diff t}{t}\, K^{(0)}(t;\sigma_0,\sigma_0)
 &=
 \frac{(1+\cM^{(0)} R^2)^{\frac{3}{2}}}{6\pi R^3}\,,\\
 \label{t-int-2}
 \int_0^\infty \frac{\diff t}{t}\, K^{(2)}(t;\sigma_0,\sigma_0)
 &=
 \frac{\sqrt{1 + \cM^{(0)} R^2}}{4\pi R} 
 \left[
 \frac{z_0^2}{6R^2} (2\tr\bar\Pi - \bar{\RicciScalar})
 + \vev{\cM^{(2)}}
 \right]\,.
\end{align}

\subsection{The bosonic determinant}
\label{sec:vector}

We can use the results above to evaluate the determinant of bosonic 
fluctuations of M2-branes describing surface operators. The quadratic 
action is derived in Appendix~\ref{app:fluct} and consists of 
eight bosonic and eight fermionic modes. The bosonic modes are 
in turn split into the four modes related to the $S^4$ part 
of the geometry and four directions transverse to the 
M2-brane world-volume within $AdS_7$.

The first four bosonic modes are massless scalars, and their kinetic operator is
simply $L = -\Delta$, see~\eqref{eqn:bosflucaction}. The path integral over these fluctations
contributes $-\frac{4}{2} \log \det(-\Delta)$ to the log of the partition
function, which can be evaluated using~\eqref{eqn:detheatkernel}. The integrand
contains both a cubic and a simple pole in $z$. Performing the $z$ integral, the
cubic pole leads to a quadratic divergence in the cutoff, which we drop (it can
be removed by a local counterterm).  The simple pole leads to $\log\epsilon$
divergence, and leaves an integral over $\bar{\Sigma}$ as in the expression for
the anomaly~\eqref{anomaly-integral}.  The contribution of these four modes to
the anomaly density is therefore 
\bal
\label{anomaly-S4}
\cA^{S^4}
&=-\frac{4}{2} \res_{z_0\to0}\left[
 \int_0^\infty \frac{\diff t}{t}
\frac{\sqrt{g}^{(-3)} + \sqrt{g}^{(-1)} + \cdots}{\sqrt{\bar{g}}} 
\left( K^{(0)}(t;\sigma_0,\sigma_0) + K^{(2)}(t;\sigma_0,\sigma_0) + \cdots \right)
\right]
\\&=
-2 \res_{z_0 \to 0} \left[
\left( -\frac{1}{2z_0} \tr{\bar{\Pi}} \right) \int_0^\infty
\frac{\diff t}{t} K^{(0)}(t;\sigma_0,\sigma_0)
+ \frac{1}{z_0^3} \int_0^\infty
\frac{\diff t}{t} K^{(2)}(t;\sigma_0,\sigma_0) + \cdots
\right]\\
&=
\frac{1}{4\pi} \frac{1}{3} \bar{\RicciScalar}\,.
\eal
The second line has the metric in Riemann normal 
coordinates~\eqref{eqn:expanded-det-metric}, and in going to the 
last line we use the expressions for the integrated heat 
kernel~\eqref{t-int-0} and~\eqref{t-int-2}.

The second set of four bosonic modes parametrise the normal 
bundle $N\Sigma$ to the world-volume of the M2-brane.
The kinetic operator derived in~\eqref{eqn:bosflucaction} acts on the fiber bundle (with
fiber $V = \bR^4$), and the path integral over the fluctuations contributes a
factor $-\frac{1}{2} \log\det_V(-\Delta + M^2)$ to the log of the partition
function. Evaluating this determinant requires a slight generalisation of the
formalism for the scalars.

The heat kernel is still defined by~\eqref{eqn:heatkerneleqn}, 
with the initial condition multiplied on the right hand side by 
the identity matrix $\mathds{1}_{V}$, and the determinant of 
$L$ has the aditional trace
\begin{align}
 \log\det\nolimits_V(L)
 =
 - \int_\Sigma \diff^3 \sigma \sqrt{g} \int_0^\infty \frac{\diff t}{t} \tr_V{K(t;\sigma,\sigma)}\,.
\end{align}

The class of differential operators we need to consider are laplacians on the
normal bundle along with a mass term. They take the form
\begin{align}
\label{vector-L}
 L = -\frac{1}{\sqrt{g}} D_{\mu} \left( \sqrt{g} g^{\mu\nu} D_\nu \right)
 + M^2\,,
\end{align}
with $D_\mu$ the covariant derivative acting on fluctuations $\zeta^{m'}$
\begin{align}
 D_\mu \zeta^{m'} \equiv
 \partial_\mu \zeta^{m'} + \omega^{m'n'}_{\mu} \zeta^{n'}\,,
 \label{eqn:normalbundleder}
\end{align}
where $\omega_\mu^{m'n'}$ is the spin connection on the 
normal bundle. To evaluate it, we pick a vielbein for the metric
$G$~\eqref{new-FG}. We then expand it at $u=0$ in $(z,\tau^a)$ as
\begin{align}
 E = E^{(-1)} + E^{(1)} + \cdots\,,
\end{align}
where $E^{(-1)}$ is a vielbein for the $AdS_7$ metric, for instance
\begin{align}
 E^{(-1)\,m}
 =
 \frac{1}{z} \delta_\mu^m \diff \sigma^\mu\,,
 \qquad
 E^{(-1)\,m'}
 =
 \frac{1}{z} \bar{E}_{\mu'}^{m'} \diff u^{\mu'}\,.
 \label{eqn:vielbeinAdS}
\end{align}

The relevant spin connection is defined by
\begin{align}
 \omega_\mu^{m'n'}
 =
 E_{\mu'}^{m'} \left( \partial_\mu E^{\mu' n'} + \Gamma^{\mu'}_{\mu \rho'}
 E^{\rho' n'}
 \right).
 \label{eqn:nbspinconnection}
\end{align}
Importantly, this vanishes using the leading order vielbein $E^{(-1)}$, 
so $\omega^{(-1)}=0$. If we expand $L$~\eqref{vector-L}, to 
leading order we recover four copies of the scalar laplacian on 
$AdS_3$~\eqref{eqn:explicitdiffopexp}. The mass matrix 
$\cM^{(0)}_{m'n'}$ is diagonal with eigenvalues $3$~\cite{Drukker:2020swu}. 
Likewise $K^{(0)}$ is a diagonal matrix with entries as in~\eqref{K0}. 
Had $\omega^{(-1)}$ not vanished, we would have had to solve for the 
vector heat kernel, in analogy with the spinors in the next section. 
Instead, the nontrivial bundle only affects $L^{(2)}$ and $K^{(2)}$.

The first subleading correction is $L^{(2)}$ as
in~\eqref{eqn:L2abstract} with the additional contribution from the spin
connection
\begin{align}
 -\frac{1}{\sqrt{g}^{(-3)}} \partial_{\mu} \left( \sqrt{g}^{(-3)}
 g^{(2)\,\mu\nu} \right) \omega_\mu^{(1)\,m'n'}
 - 2 g^{(2)\,\mu\nu} \omega^{(1)\,m'n'}_\mu \partial_\nu\,.
\end{align}
Here $\omega^{(1)}$ is the subleading correction to the spin connection. 
$K^{(2)}$ is then given by~\eqref{eqn:K2int}, and we are interested 
in the trace of the heat kernel
$\tr_V{K^{(2)}}$. Using that $\cM^{(0)}$ is a diagonal matrix (and so is
$K^{(0)}$), while $\omega^{(1)}$ is antisymmetric, we see that these
additional contributions simply drop out of the trace
\begin{align}
 \tr_V{K^{(0)} \omega^{(1)} K^{(0)}} = 0\,.
\end{align}

We therefore do not even need to write down explicit expressions 
for the corrections $E^{(1)}$ and $\omega_\mu^{(1)\,m'n'}$. 
Instead, $K^{(2)}$ is a straightforward generalisation of the 
scalar case, where the only nontrivial matrix structure is in 
$\cM^{(2)}$. The result is the same as~\eqref{eqn:k2scalarres}
\begin{align}
 \tr_V{K^{(2)}(t;\sigma_0,\sigma_0)}
 =
 \tr_V\left[ \left(
 \frac{z_0^2}{6}(\bar{\RicciScalar} - 2 \tr\bar\Pi)
 - \vev{\cM^{(2)}}
 \right)
 \frac{\exp\left( -(1+\cM^{(0)})t \right)}
 {(4 \pi)^{3/2} t^{1/2}} \right]\,.
 \label{eqn:K2coincidentvec}
\end{align}

For the M2-brane, the mass matrix $M^2$ is a function over the 
world-volume $\Sigma$ derived in~\eqref{eqn:bosflucaction}. 
Its trace over the four transverse $AdS_7$ modes is
\begin{align}
 M^2_{m'm'}
 =
 - g^{\mu\nu} G^{\mu'\nu'} R_{\mu' \mu \nu' \nu}
 - g^{\mu \nu} g^{\rho \sigma} G_{\mu'\nu'} \tilde{\IIFundForm}_{\mu\rho}^{\mu'}
 \tilde{\IIFundForm}_{\sigma \nu}^{\nu'}\,,
\end{align}
where here $R_{\mu'\mu\nu'\nu}$ is the bulk Riemann tensor (as opposed to the
world-volume Riemann tensor $\cR$).
In order to read $\cM^{(0)}$ and $\cM^{(2)}$ we rewrite $M^2$ as follows.
Recall that $\mu$, $\nu$ are the M2-brane directions and $\mu'$, $\nu'$ 
the transverse ones, so we can write
\begin{align}
 G^{\mu'\nu'} R_{\mu' \mu \nu' \nu}
 =
 R_{\mu\nu} - g^{\rho \sigma} R_{\mu \rho \nu \sigma}\,.
\end{align}
The Ricci tensor is fixed by the bulk equations of motion
\begin{align}
 R_{\mu\nu} = -6 G_{\mu\nu} = - 6 g_{\mu\nu}\,,
\end{align}
and for the components of the Riemann tensor, we relate them 
to the world-volume curvature $\cR$ via the Gauss-Codazzi equation
\begin{align}
 R_{\mu \rho \nu \sigma} = \cR_{\mu \rho \nu \sigma}
 + 2 \IIFundForm^{\mu'}_{\mu [\sigma}
 \IIFundForm^{\nu'}_{\nu] \rho} G_{\mu'\nu'}\,.
\end{align}
Together we then find
\begin{align}
 M_{m'm'}^2
 =
 18 + \RicciScalar\,,
\end{align}
which is now expressed solely in terms of quantities depending on the induced metric
$g$~\eqref{asym-met-3}. It is easy to show that the
world-volume Ricci scalar has expansion $\RicciScalar= -6 +
\RicciScalar^{(2)} + \cdots$ with
\begin{align}
 \RicciScalar^{(2)}
 =
 \frac{z_0^2}{\left( \cosh \rho - \cos \theta \sinh \rho \right)^2} (\bar{\RicciScalar} - 2 \tr
 \bar{\Pi})\,.
 \label{eqn:ricciscalarcorr}
\end{align}
The leading trace of the mass matrix is $\tr{\cM^{(0)}} = 12$, as expected for
four modes of mass-squared 3. The subleading correction is
\begin{align}
 \tr{\cM^{(2)}}
 =\RicciScalar^{(2)}
 =\frac{z_0^2}{(\cosh{\rho} - \cos{\theta} \sinh{\rho})^2}
 (\bar{\RicciScalar} - 2 \tr{\bar{\Pi}})\,.
\end{align}
Integrating over $S^2$,
\begin{align}
 \vev{\tr{\cM^{(2)}}}
 &=
 \frac{1}{4\pi} \int \diff \theta\, \diff \phi \sin{\theta}
 \tr{\cM^{(2)}}
 =
 z_0^2 \left( \bar{\RicciScalar} - 2 \tr{\bar{\Pi}}
 \right)\,.
 \label{eqn:tracem2average}
\end{align}
As we anticipated in~\eqref{eqn:k2Ricciint}, this does not depend on $\rho$.

To get the contribution to the anomaly, we again take the $t$ 
integral of the trace of the heat kernel~\eqref{t-int-0}, 
\eqref{t-int-2}, and look for the residue at $z_0\to0$ as 
in \eqref{anomaly-S4}. We get
\bal
\label{anomaly-NS}
\cA^{N\Sigma}
&=-\frac{1}{2}
\res_{z_0 \to 0} \left[
\left( -\frac{1}{2z_0} \tr{\bar{\Pi}} \right) \int_0^\infty
\frac{\diff t}{t} \tr_V K^{(0)}(t;\sigma_0,\sigma_0)
+ \frac{1}{z_0^3} \int_0^\infty
\frac{\diff t}{t} \tr_V K^{(2)}(t;\sigma_0,\sigma_0) + \cdots
\right]\\
&=
\frac{1}{4\pi}\left( -\frac{1}{3} \bar{\RicciScalar} + 6 \tr{\bar{\Pi}} \right).
\eal

\subsection{Spinor bundles}
\label{sec:spinors}

A similar analysis can be performed for spinors. The Dirac operator in euclidean
signature is $\slashed{D} - M$ and it acts on the spinor bundle $S$ with fiber
$\bC^8$ corresponding to the (chiral) spinor representations of
$SO(3)$, $SO(4)_N$ for normal directions and $SO(4)_S$ for $S^4$ directions.
We are interested in the determinant of its square
\begin{align}
 \det\left( -\slashed{D}^2 + M^2 \right)
 \equiv
 \det\left( L_F \right)\,.
\end{align}
The square of the Dirac operator is related to the spinor laplacian
via a generalisation of the Lichnerowicz formula (see
e.g.~\cite{lawson}). Using $\gamma_m$ and $\rho_{m'}$ for the gamma matrices of
$SO(3)$ and $SO(4)_N$ respectively 
(so $\left\{ \gamma_m, \gamma_n \right\} = 2\delta_{mn}$ 
and similarly for $\rho$) and $\gamma_{mn} = \frac{1}{2} \left[
\gamma_m, \gamma_n \right]$ for the antisymmetrised product, it reads
\begin{align}
\label{D-slash2}
 \slashed{D}^2
 =
 \Delta_{1/2} - \frac{\RicciScalar}{4}
 + \frac{1}{8} \gamma_{m n} \rho_{m'n'} R_{m n m'n'}\,.
\end{align}
$\Delta_{1/2}$ is spinor laplacian given in terms of the spinor covariant
derivative as
\begin{align}
 \Delta_{1/2}
 =
 \frac{1}{\sqrt{g}} D_\mu \left( \sqrt{g}g^{\mu\nu} D_\nu \right)\,,
 \qquad
 D_\mu \psi
 =
 \left( \partial_\mu - \frac{1}{4} \gamma_{mn} \omega_{\mu}^{mn} - \frac{1}{4}
 \rho_{m'n'} \omega_{\mu}^{m'n'} \right) \psi\,,
\end{align}
and $R$ is the curvature 2-form of the normal bundle
\begin{align}
 R_{\mu \nu}^{m'n'} = 2 \partial_{[\mu} \omega_{\nu]}^{m'n'} + 2
 \omega_{[\mu}^{m'p'} \omega_{\nu]}^{p'n'}\,.
\end{align}
Note that we assume here again that the embedding in $S^4$ is trivial, 
so we do not see its spin connection.

To proceed, we again look for a near-$AdS_3$ expansion
of the operator $L_F = -\slashed{D}^2 + M^2$ and the corresponding heat kernel. 
We also restrict to constant $M^2 = \cM^{(0)}$. We then require the generalisation of
Section~\ref{sec:kernel} to the case of spinor bundles on $AdS_3$. 

To leading order, the metric~\eqref{expanded-metric} is $AdS_3$ and the
differential operator $L^{(0)}$ is the $AdS_3$ spinor laplacian. 
The spinor heat kernel derived in~\cite{Camporesi:1992tm} (see
also~\cite{Camporesi:1995fb,David:2009xg}) 
transforms nontrivially as a bispinor at points $\sigma$ and
$\sigma_0$. It takes the factorised form
\begin{align}
 K^{(0)}(t;\sigma,\sigma_0)
 &=
 U(\sigma, \sigma_0) k_F(t;\rho)\,,
\end{align}
where $U$ encodes the bispinor transformations and $k_F$ is a function of $t$
and the geodesic distance $\rho$ between $\sigma$ and $\sigma_0$ only. On a curved background, $U$ can
be obtained by parallel transporting a spinor along the geodesic connecting
$\sigma$ and $\sigma_0$
\begin{align}
\label{U-general}
U(\sigma, \sigma_0) = \mathcal{P} \exp{\int_{\sigma_0}^{\sigma} 
 \diff x^\mu\, \frac{1}{2}
 \omega_\mu^{mn} \gamma_{mn}}\,.
\end{align}

To write explicit expressions, we use the coordinates in~\eqref{eqn:AdS3} 
with $\sigma_0$ the origin of $AdS_3$ and the vielbein
\begin{equation}
\begin{aligned}
 e^1 &= \sin\theta \cos \phi\,\diff \rho + \sinh{\rho}\left( \cos{\theta}
 \cos{\phi}\,\diff \theta - \sin{\theta} \sin{\phi}\,\diff \phi
 \right)\,,\\
 e^2 &= \sin{\theta} \sin{\phi}\,\diff \rho + \sinh{\rho} \left( \cos{\theta}
 \sin{\phi}\,\diff \theta + \sin{\theta} \cos{\phi}\,\diff \phi
 \right)\,,\\
 e^3 &= \cos{\theta}\,\diff \rho - \sinh{\rho} \sin{\theta}\,\diff \theta\,.
\end{aligned}
\label{eqn:parallelframe}
\end{equation}
This frame is diagonal in projective 
coordinates and the corresponding spin connection is
\begin{equation}
\begin{aligned}
 \omega^{12} &=
 -2 \sinh^2{\frac{\rho}{2}} \sin^2{\theta}\,\diff \phi\,,\\
 \omega^{13} &=
 2 \sinh^2{\frac{\rho}{2}} \left( \cos{\phi}\,\diff\theta - \cos{\theta} \sin{\theta}
 \sin{\phi}\,\diff \phi \right),\\
 \omega^{23} &=
 2 \sinh^2{\frac{\rho}{2}} \left( \sin{\phi}\,\diff\theta + \cos{\theta} \sin{\theta}
 \cos{\phi}\,\diff\phi \right).
\end{aligned}
\end{equation}
Now the geodesic connecting the origin to a point 
$\sigma = (\rho, \theta, \phi)$ is along
the vector $\partial_\rho$. Since the spin connection along
that direction is trivial $\omega_\rho^{mn} = 0$, this 
frame is known as a parallel frame and the matrix 
$U$~\eqref{U-general} reduces to the identity matrix%
\footnote{%
Note that also for the frame 
$e=(\diff\rho,\sinh\rho\,\diff\theta,\sinh\rho\sin\theta\,\diff\phi)$ 
the spin connection component $\omega_\rho^{mn}=0$, so $U$ is 
independent of $\rho$. However, the spin connection does not vanish at 
$\rho=0$, which introduces angular dependence into $U$. It can still be 
computed, as was done in the case of $AdS_2$ in \cite{Bergamin:2015vxa}.}
\begin{align}
 U(\sigma, \sigma_0) 
 = \mathds{1}_8\,.
\end{align}
Using this frame, the heat kernel equation reduces to
\begin{equation}
\begin{aligned}
 &(L_F^{(0)}+\partial_t) K^{(0)}(t;\sigma,\sigma_0)
 =
 \left[-\partial_\rho^2 - 2 \coth{\rho} \,\partial_\rho + \frac{1}{2} \tanh^2\frac{\rho}{2}
 - \frac{3}{2} + \cM^{(0)} + \partial_t \right] k_F(t;\rho) =
 0\,,
\end{aligned}
\label{eqn:heatkernleqnspinor}
\end{equation}
where $\frac{1}{2} \tanh^2({\rho}/{2})$ is the leading order
contribution of $g^{\mu\nu} \omega_\mu^{mn} \omega_{\nu}^{mn}/8$ and
$-{3}/{2} ={\RicciScalar^{(0)}}/{4}$ is the scalar
curvature of $AdS_3$.
If we define
\beq
\label{kFh}
k_F(t;\rho)
=-\frac{1}{4\pi\sinh(\rho/2)}\partial_\rho\left(\frac{h(t;\rho)}{\cosh(\rho/2)}\right),
\eeq
then $h$ satisfies the one-dimensional heat kernel equation
\beq
\label{1d}
\left[-\partial_\rho^2 + \cM^{(0)} + \partial_t \right] h(t;\rho) =
 0\,.
\eeq
A solution to this equation is
\beq
 h(t;\rho)
 =
 \frac{\exp{\left( -\cM^{(0)} t - {\rho^2}/{4t} \right)}}
 {\sqrt{4\pi t}}\,,
\eeq
and its derivative 
\begin{align}
 k_F(t;\rho)
 &=
 \frac{\rho + t \tanh({\rho/2})}{\sinh{\rho}}
 \frac{\exp{\left( -\cM^{(0)} t - {\rho^2}/{4t} \right)}}{(4\pi t)^{3/2}}\,,
 \label{eqn:heatkernelspinorsol}
\end{align}
satisfies the initial 
condition~\eqref{eqn:heatkerneleqn}~\cite{Camporesi:1992tm}.

To find the subleading correction to the heat kernel $K^{(2)}$, we can again
use~\eqref{eqn:K2int}, which extends to the spinor case. One obvious 
difference is that the differential operator \eqref{D-slash2} contains 
extra terms compared to the scalar laplacian
\begin{equation}
\begin{aligned}
 L_F^{(2)}
 =
 \left[ 
 L^{(2)}
 +
 \frac{1}{8} g^{(4)\,\mu\nu} \omega_\mu^{(-1)\,mn}
 \omega_{\nu}^{(-1)\,mn} 
 + \frac{1}{4} g^{(2)\,\mu\nu}\omega_\mu^{(1)\,mn}
 \omega_{\nu}^{(-1)\,mn}
 + \frac{\RicciScalar^{(2)}}{4}
 \right] \mathds{1}_8 + \cdots
\end{aligned}
 \label{eqn:L2spinor}
\end{equation}
where the ellipses involve terms with single gamma matrices which vanish upon 
taking the trace over the spinor bundle. 
One should also worry about the matrix $U$. As stated, 
in our frame $U(\sigma',\sigma_0)=\mathds{1}$, but the same is not generally 
true for $U(\sigma,\sigma')$. As we only require $K^{(2)}(\sigma_0,\sigma_0)$, 
the second heat kernel comes with $U(\sigma_0,\sigma')$, which is also the 
identity.

The correction to the metric in \eqref{eqn:L2spinor} 
is given in~\eqref{expanded-metric} and that of the
scalar curvature is~\eqref{eqn:ricciscalarcorr}.
To obtain that of the spin connection, use that the vielbein for the
asymptotically $AdS_3$ metric~\eqref{expanded-metric}
is
\begin{align}
 e^{m}_\mu
 =
 e^{(-1)\,m}_{\mu} + \frac{1}{2} g_{\mu\nu}^{(0)} e^{(1)\,\nu}_m + \cdots
\end{align}
with $e^{(-1)\,m}_\mu$ the leading vielbein defined in~\eqref{eqn:parallelframe} 
and $e^{(1)\,\mu}_m$ its inverse. 
Note that $g^{(0)}$ can be written in the
$\rho, \theta, \phi$ coordinates using the change of
variables~\eqref{eqn:ads3changeofvariables}.

The expression for $K^{(2)}$ is then as in~\eqref{eqn:K2int}. Explicitly, 
\begin{align}
-\int_0^t \diff t' \int_\Sigma \diff \rho \,\diff \theta \,\diff \phi
 \sinh^2{\rho} \sin{\theta}\,
 k_F(t - t';\sigma_0,\sigma') \tr_S L_F^{(2)}
 k_F(t';\sigma',\sigma_0)\,.
\end{align}

We then need to do the spherical integral, as in \eqref{eqn:integratedL2scalar}. 
For the scalar laplacian, we get the same as above and for the other terms 
in~\eqref{eqn:L2spinor} we find
\begin{align}
 \frac{1}{8} \int \sin{\theta} \,\diff \theta\, \diff \phi\,
 g^{(4)\,\mu\nu} \omega_\mu^{(-1)\,mn} \omega_{\nu}^{(-1)\,mn}
 &=
 \frac{2 \pi}{3} \bar{\Pi}\tanh^2\frac{\rho}{2}
 \label{eqn:corrspincon1}
 + \frac{\pi}{6} \bar{\RicciScalar}\left(\rho \cosh{\rho} -
 \sinh{\rho}\right)\frac{\sinh(\rho/2)}{\cosh^3(\rho/2)}\,,
 \\
 \frac{1}{4} \int \sin{\theta} \,\diff \theta \,\diff \phi\,
 g^{(2)\,\mu\nu} \omega_\mu^{(1)\,mn} \omega_{\nu}^{(1)\,mn}
 &=
 -\frac{\pi}{3}\bar{\RicciScalar} \rho \tanh{\frac{\rho}{2}} \,,
 \label{eqn:corrspincon2}\\
 \frac{1}{4} \int \sin{\theta} \,\diff \theta\, \diff \phi\,
 \RicciScalar^{(2)}
 &=
 \pi \left(\bar{\RicciScalar} - 2 \bar{\Pi} \right)\,.
 \label{eqn:corrricciint}
\end{align}

Assembling the results, we obtain
\begin{equation}
\begin{aligned}
 -4\pi
 & \int_0^t \diff t' \int_0^\infty \diff \rho \sinh^2{\rho}\,
 k_F(t-t';\rho) 
 \left[
 \frac{z_0^2}{3}\tr\bar\Pi
 \left( -\partial_\rho^2 - 2 \coth{\rho}\,\partial_\rho + \frac{1}{2} \tanh^2\frac{\rho}{2} -
 \frac{3}{2} \right) \right.\\
 & \qquad \left.
 + z_0^2 \bar{\RicciScalar}
 \left(
 \frac{1}{6} \frac{\cosh{\rho} \sinh{\rho} - \rho}{\sinh^2\rho} \partial_\rho
 + \frac{1}{4} - \frac{\rho + \sinh{\rho}}{24}
 \frac{\sinh(\rho/2)}{\cosh^3(\rho/2)}
 \right)
 \right]
 k_F(t';\rho)
 \tr_S(\mathds{1}) \,.
 \label{eqn:spinorheatkernelrhointegral}
\end{aligned}
\end{equation}
This integral is evaluated explicitly in Appendix~\ref{app:integral}. 
%
Adding~\eqref{eqn:k2ftrPiint} and \eqref{eqn:k2fricciint2} together with the
boundary term appearing in the second line of~\eqref{eqn:K2int}, we finally get
the trace of the heat kernel
\begin{align}
 \tr_S K^{(2)}(t;\sigma_0,\sigma_0)
 &=
 - \frac{z_0^2}{12} (\bar{\RicciScalar} - 2 \tr{\bar{\Pi}})
 \frac{e^{-\cM^{(0)} t}}{(4\pi)^{3/2} t^{1/2}} \tr_S{\mathds{(1)}}\,.
\end{align}
Its $t$ integral is easy to evaluate and gives (reinstating powers of $R$)
\begin{align}
 \label{eqn:K0ftint}
 \int_0^\infty \frac{\diff t}{t} \tr_S K^{(0)}(t;\sigma_0,\sigma_0)
 &=
 \frac{\sqrt{\cM^{(0)}}}{24\pi R^2} (4 \cM^{(0)} R^2 - 3) \tr_S(\mathds{1}) \,,\\
 \label{eqn:K2ftint}
 \int_0^\infty \frac{\diff t}{t} \tr_S K^{(2)}(t;\sigma_0,\sigma_0)
 &=
 \frac{z_0^2}{48\pi R^2} (\bar{\RicciScalar} - 2 \tr{\bar{\Pi}})
 \sqrt{\cM^{(0)}} \tr_S{(\mathds{1})} \,.
\end{align}

Getting from this to the contribution of the fermions to the anomaly 
of the surface operator is immediate. The Dirac operator governing the
quadratic fermionic fluctuations for an M2-brane is derived in
Appendix~\ref{app:fluct}, and from equation~\eqref{eqn:fermionic-det} we can 
read $\cM^{(0)} =9/4$.

Evaluating the path integral yields the pfaffian of the Dirac operator, or
$\frac{1}{2} \log \det_S(- \slashed{D}^2 + 9/4)$. The contribution of
the determinant to the anomaly is obtained from the pole of the $z$ integral as
in the bosonic case~\eqref{anomaly-S4}. Using~\eqref{eqn:K0ftint}, \eqref{eqn:K2ftint}, 
and $\tr_S(\mathds{1})=8$ we get
\begin{align}
 \nonumber \cA^{F}
 &=
 \frac{1}{2}
 \res_{z_0 \to 0} \left[
 \left( -\frac{1}{2z_0} \tr{\bar{\Pi}} \right) \int_0^\infty
 \frac{\diff t}{t} \tr_S K^{(0)}(t;\sigma_0,\sigma_0)
 + \frac{1}{z_0^3} \int_0^\infty
 \frac{\diff t}{t} \tr_S K^{(2)}(t;\sigma_0,\sigma_0) + \cdots
 \right]\\
 &=
 \frac{1}{4\pi} \left(
 \frac{1}{2} \bar{\RicciScalar} - 4 \tr{\bar{\Pi}} \right).
 \label{anomaly-fermion}
\end{align}

\section{Conclusions}
\label{sec:conclude}

Combining the results of the massless scalars on $S^4$ \eqref{anomaly-S4}, the 
normal bundle inside $AdS_7$ \eqref{anomaly-NS}, the 
fermions \eqref{anomaly-fermion}, and replacing $\tr\bar\Pi=H^2/4+\tr\bar P$ 
\eqref{Pi}, we find
\begin{align}
 \frac{1}{2} \log 
 \frac{\det_S{\left(-\slashed{D}^2+9/(4R^2)\right)}}
 {\det^4(-\Delta) \det_V(-\Delta_{m'n'} + M^2_{m'n'})}
 &=
 \frac{\log{\epsilon}}{4\pi} \int_{\bar{\Sigma}} \vol_{\bar{\Sigma}} \left[
 \frac{1}{2} \bar{\RicciScalar} + \frac{1}{2} (\bar{H}^2 + 4 \tr{\bar{P}})
 \right] + \cdots
\end{align}
where the $\cdots$ stand for quadratic divergences in $\epsilon$ and finite
terms.  The integrand takes the form of an anomaly density \eqref{eqn:anomaly}
and from it we read the order $N^0$ corrections to the anomaly coefficients,
respectively $1/2$, $1/2$ and 0 for $\bar{\cR}$, $\bar{H}^2 + 4 \tr{\bar{P}}$,
$\tr{\bar{W}}$, giving \eqref{coefficients}. This provides a first-principles
evaluation of these anomaly coefficients previously derived in~\cite{
Jensen:2018rxu, 
Rodgers:2018mvq,Estes:2018tnu, Wang:2020xkc, 
Bullimore:2014upa,Chalabi:2020iie,Meneghelli:2022gps, 
Bianchi:2019sxz, Drukker:2020atp, Drukker:2020swu}.

While the values of the anomaly coefficents are already known, the tools we
develop are new. In particular, we obtain explicit expressions for the trace of
the heat kernel associated to massive scalars~\eqref{t-int-2} and
spinors~\eqref{eqn:K2ftint} in asymptotically $AdS_3$ spaces. Using these
results we are able to extend the classical near boundary asymptotic analysis of
Graham-Witten~\cite{graham:1999pm} to the quantum level.


Along the way we also obtain the quadratic fluctuation action about any
classical M2-brane solution in $AdS_7\times S^4$, which is presented in
Appendix~\ref{app:fluct}. 
These expressions reproduce known results when specifying the appropriate
classical solution, in particular we recover the spectrum of fluctuations of the
parallel planes~\cite{Forste:1999yj} and sphere~\cite{Drukker:2020swu}.
In these cases and the other few examples of known classical M2-brane
solution~\cite{mezei:2018url, Drukker:2021vyx, Drukker:2022beq}, our results
provide the determinants capturing the 1-loop fluctuations around the classical
solutions. One may now try to evaluate these determinants exactly and go beyond 
the analysis of divergences that lead to the anomalies studied here.

Our work also serves to enlarge the scope of recent progress on the 
semiclassical quantisation of M2-branes. This has been 
done in several contexts like Wilson loops and 
instanton corrections to the free energy of 
ABJM theory~\cite{Giombi:2023vzu,Beccaria:2023ujc, Seibold:2023zkz, 
Beccaria:2023sph, toappear} and reproduced results from 
supersymmetric localisation. In most of those cases the 
classical M2-brane solutions are BPS and their spectrum can 
be solved exactly and summed up using $\zeta$ function regularisation. 
Here the brane does not need to preserve supersymmetry.

One limitation of our calculation is that we assume that 
the classical surface is located at a single point in $S^4$, 
which is not the most general setting (and in particular not sufficient for most
of the cases studied in~\cite{Drukker:2020bes, Drukker:2021vyx, Drukker:2022beq}). 
When the classical M2-brane is extended in $S^4$, there is an extra anomaly coefficient 
denoted $c$ in \cite{Drukker:2020dcz}---if the embedding into 
$S^4$ is represented by a unit vector $n^i(\tau^a)$, then 
$c$ multiplies $(\partial n^i)^2$. This term can be thought 
more generally as arising from the coupling of a surface operator 
to a global symmetry. It was proven in~\cite{Drukker:2020atp} 
that in the ${\cal N}=(2,0)$ theory $c=-a_2$, so our 
analysis here establishes that it is $c=N-1/2+\cO(1/N)$. 
Nonetheless, it would be instructive to rederive it from 
first principles, by generalising the analysis in this paper 
to generic embeddings in $AdS_7\times S^4$.

Finally, our analysis suggests many further directions. Perhaps the
most straightforward is the calculation of anomaly coefficients for the
non-supersymmetric surface operator of the 6d $\cN=(2,0)$
theory~\cite{Drukker:2020swu}. This is the surface operator analog of the 
usual Wilson loop in ${\cal N}=4$ supersymmetric Yang-Mills without scalar 
coupling~\cite{Polchinski:2011im,Beccaria:2017rbe}, and in holography
its scalar fluctuations on $S^4$ obey Neumann instead of 
Dirichlet boundary conditions; we hope to report on it soon.
It should also be possible to calculate anomaly coefficients for surface operators in
large representations using a probe M5-brane as 
in~\cite{Chen:2007ir, Mori:2014tca} and its fluctuations.
More generally, a similar analysis should also be possible for other settings
where the asymptotic analysis proved useful,
for instance in the calculation 
of entanglement
entropy~\cite{Solodukhin:2008dh,Lewkowycz:2014jia,Bianchi:2015liz},
the study of surface operators in 4d~\cite{Gukov:2006jk, Drukker:2008wr} 
and that of codimension-2 observables in 
6d~\cite{Chalabi:2021jud,Capuozzo:2023fll}.

\subsection*{Acknowledgements}

We are grateful to C.R. Graham, F. H\"ubner, L. Pando Zayas, G. Papadopoulos, R. Sinha, A.
Tseytlin and E. Witten for helpful discussions, and especially A. Tseytlin his
many helpful comments on a preliminary version of this paper.  ND would like to
thank EPFL, CERN and DESY for their hospitality in the course of this work. ND’s
research is supported by the Science Technology \& Facilities council under the
grants ST/P000258/1 and ST/X000753/1.  OS's research is funded by the
Engineering \& Physical Sciences Research Council under grant number
EP/W524025/1.  MT gratefully acknowledges the support from the Institute for
Theoretical and Mathematical Physics (ITMP, Moscow) where this project began,
and the Perimeter Institute where part of this project was realised.  MT’s
research is funded by the Engineering \& Physical Sciences Research Council
under the grant EP/W522429/1. 

\appendix
\section{Quadratic M2-brane action}
\label{app:fluct}

In this appendix we derive the action for quadratic fluctuations of 
M2-branes in $AdS_7\times S^4$. This is used in Section~\ref{sec:det} 
where the determinant of the kinetic operators are evaluated. 
The spectrum of fluctuations of specific M2-brane classical solutions 
in different backgrounds have been previously calculated 
in~\cite{Forste:1999yj,Giombi:2023vzu, Beccaria:2023ujc, 
Beccaria:2023sph, toappear}. The calculation follows closely 
that of the fluctuations of semiclassical strings pioneered 
in~\cite{Drukker:2000ep}, elaborated on in~\cite{Forini:2015mca} 
and generalised to other branes in~\cite{deLeonArdon:2020crs}.

For the purpose of this paper it would be enough to study the 
asymptotic form of the M2-brane quadratic fluctuation action, 
but it turns out to be simple enough to derive it for 
any classical solution. In this appendix we therefore 
abandon the near boundary metric and use one near the classical brane solution.
Indeed, the brane does not need to have a boundary 
and can be a bulk instanton, as those studied 
in~\cite{Beccaria:2023ujc,Beccaria:2023sph}.

As in Section~\ref{sec:3metrics}, we take
$\sigma^\mu$ as the three world-volume coordinates 
parametrising $\Sigma$ and for the remaining eight directions, we take $u^{\mu'}$ 
to be normal to $\Sigma$ and $z^i$ to parametrise $S^4$, such that the
classical solution is at $u^{\mu'}=0$ and $z^i = 0$.
We ignore global issues and write the metric near the brane as
\begin{align}
\label{metric-split}
 \diff s^2
 =
 G_{\mu \nu}(\sigma,u) \diff \sigma^\mu \diff \sigma^\nu
 + 2 G_{\mu \mu'}(\sigma,u) \diff \sigma^\mu \diff u^{\mu'}
 + G_{\mu' \nu'}(\sigma,u) \diff u^{\mu'} \diff u^{\nu'}
 + \frac{\diff z^i \diff z^i}{(1+z^j z^j)^2}\,,
\end{align}
with $G_{\mu\nu}(\sigma,0) = g_{\mu\nu}(\sigma)$ the induced metric on $\Sigma$ 
and $G_{\mu \mu'}(\sigma,0)= 0$. Note that the metric in 
\eqref{new-FG} satisfies all the same properties, so realises 
this in the near-boundary regime.

We proceed to analyse the bosonic and fermionic fields on the 
M2-brane world-volume in the geometry \eqref{metric-split} in turn.

\subsection{Bosonic fluctuations}
\label{sec:bosonic}

We consider fluctuations $u^{\mu'}, z^i$ around a classical solution 
described by world-volume coordinates $x^\mu$ as in the metric 
\eqref{metric-split}. 
The M2-brane action \eqref{action} contains two terms, the volume form and a 
term coupling to the pullback of the three-form $A_3$.
Because we restrict to classical solutions localised at a point on $S^4$, there
are no tangent vectors along $S^4$ and the pullback of $A_3$ is at least cubic
in the fluctuations, and we can neglect it.
The quadratic terms arising from the volume form 
is obtained in~\cite{deLeonArdon:2020crs}%
\footnote{But correcting a factor of 2 in~\cite{deLeonArdon:2020crs}.}
(see also~\cite{Goon:2020myi})
\begin{align}
 \frac{T_\text{M2}}{2} \int_\Sigma \vol_\Sigma \left[ 
 \partial_\mu u^{\mu'} \partial_\nu u^{\nu'} g^{\mu\nu} g_{\mu' \nu'}
 + \partial_\mu z^i \partial_\nu z^i g^{\mu\nu}
 - \left( R_{\mu' \mu \nu' \nu}
 + g_{\mu' \rho'} g_{\nu' \sigma'} \tilde{\IIFundForm}^{\rho'}_{\mu\rho}
 \tilde{\IIFundForm}^{\sigma'}_{\sigma \nu} g^{\rho \sigma}
 \right) g^{\mu\nu}
u^{\mu'} u^{\nu'} \right],
\label{eqn:m2bosflucu}
\end{align}
with $\tilde{\IIFundForm}$ the traceless part of the second fundamental form and
$R$ the Riemann tensor. 

To put the kinetic term in canonical form, we 
introduce the vielbeine $E^{\mu'}_{m'}(\sigma)$ 
and parametrise $u$ as
\begin{align}
 u^{\mu'} = E^{\mu'}_{m'}\zeta^{m'}\,.
\end{align}
The derivatives are then expressed as
\begin{align}
 \partial_\mu u^{\mu'} = E^{\mu'}_{m'} D_\mu \zeta^{m'}\,,
 \qquad
 D_\mu \zeta^{m'} \equiv
 \partial_\mu \zeta^{m'} + \omega^{m'n'}_{\mu} \zeta^{n'}\,,
 \label{eqn:defncovdernormal}
\end{align}
with $\omega$ the spin connection on the normal bundle.
With these substitutions the quadratic action becomes
\begin{align}
 \frac{T_\text{M2}}{2} 
 \int_\Sigma \vol_\Sigma \left[ 
 D_\mu \zeta^{m'} D_\nu \zeta^{m'} g^{\mu\nu}
 - \left(
 E^{\mu'}_{m'} E^{\nu'}_{n'} R_{\mu' \mu \nu' \nu}
 + E_{\mu'm'} E_{\nu'n'} \tilde{\IIFundForm}^{\mu'}_{\mu\rho}
 \tilde{\IIFundForm}^{\nu'}_{\sigma \nu} g^{\rho \sigma}
 \right) g^{\mu\nu} \zeta^{m'} \zeta^{n'}
 \right].
\end{align}

Finally, to bring this to the form of a differential operator, we integrate the
kinetic term by parts. This yields a boundary term, and assuming it vanishes, 
we get \cite{deLeonArdon:2020crs}
\begin{equation}
\begin{aligned}
 S_\text{fluc}^\text{(bos)} &= 
 \frac{T_\text{M2}}{2} \int_\Sigma \vol_\Sigma \zeta^{m'}
 \left[ - \Delta_{m'n'} + M^2_{m'n'} \right] \zeta^{n'} - z^i \Delta z^i \,,\\
 M^2_{m'n'} &=
 -g^{\mu\nu} \left( 
 E^{\mu'}_{m'} E^{\nu'}_{n'} R_{\mu' \mu \nu' \nu}
 + E_{\mu'm'} E_{\nu'n'} \tilde{\IIFundForm}^{\mu'}_{\mu\rho}
 \tilde{\IIFundForm}^{\nu'}_{\sigma \nu} g^{\rho \sigma}\right),
\end{aligned}
 \label{eqn:bosflucaction}
\end{equation}
with $\Delta$ the usual laplacian and $\Delta_{m'n'}$ the vector laplacian acting
on $\zeta^{m'}$ (the covariant derivative is defined
in~\eqref{eqn:defncovdernormal})
\beq
\label{vec-lap}
 \Delta_{m'n'}\zeta^{n'}
 = \frac{1}{\sqrt{g}} \left[D_{\mu} \left( \sqrt{g} g^{\mu\nu} D_\nu
 \right)\zeta\right]_{m'}\,.
\eeq

\subsection{Fermionic fluctuations}
\label{sec:fermionic}

The fermionic part of the M2-brane action is obtained
in~\cite{Bergshoeff:1987cm} and is expressed in terms of an 11d spinor
$\Psi(\sigma)$ (whose indices we suppress). On $AdS_7 \times S^4$
$\Psi$ is in the spinor representation of $\sof(1,6)$ and $\sof(4)$. 
To quadratic order and working in lorentzian signature, it
reads~\cite{deWit:1998tk,deWit:1998yu} (see also~\cite{Sakaguchi:2010dg})
\begin{align}
 T_\text{M2}
 \int_\Sigma \diff^3 \sigma\sqrt{-g}
 \left[ \bar{\Psi} \left(
   g^{\rho \mu} \Gamma_\mu
   - \frac{1}{2} \varepsilon^{\mu\nu\rho} \Gamma_{\mu\nu} \right)
   \left(D_\rho + \frac{1}{288} \left( {\Gamma_\rho}^{NPQR} - 8
 \delta_\rho^{[N} \Gamma^{PQR]} \right) F_{NPQR} \right) \Psi
 \right] \,.
 \label{eqn:m2actionf}
\end{align}
Here we use $M,N,\cdots$ for coordinates on the full 11d space.
$\Gamma^M$ are the 11d gamma matrices and we use the shorthand for
antisymmetrised indices $\Gamma_{\mu \nu} \equiv \Gamma_{[\mu} \Gamma_{\nu]}$ 
(with the appropriate 1/2).
The spinor covariant derivative is
\begin{align}
 D_\rho \Psi = \partial_\rho \Psi
 - \frac{1}{4} \omega_\rho^{MN} \Gamma_{MN} \Psi
 \,,
 \label{eqn:defnspinorcovder}
\end{align}
where to avoid introducing extra notations, 
we use curved space indices on the spin connection and gamma matrices.

As mentioned previously, we restrict to M2-branes located at a point on $S^4$,
so the term proportional to $\delta_\rho^{[N} \Gamma^{PQR]}$
vanishes and the covariant derivative~\eqref{eqn:defnspinorcovder} includes only
the $AdS_7$ spin connection. We can also simplify the remaining flux term by
noting that $\Gamma_\rho$ commutes with $\Gamma^{NPQR}$ and using 
the explicit expression for the flux~\eqref{eqn:f4} we get
\begin{align}
 \frac{1}{288} \Gamma_{\rho} \Gamma^{NPQR} F_{NPQR}
 =
 \frac{1}{2R} \Gamma_{\rho} \Gamma_{789 \natural}\,.
\end{align}

To proceed, we can pick a frame where 
$E_\mu^{m'} = E_{\mu'}^m = 0$. So the gamma matrices used above 
can now be expressed in terms of those with flat space indices as 
$\Gamma_\mu = E_\mu^m \Gamma_m$ 
and $\Gamma_{\mu'} = E_{\mu'}^{m'}\Gamma_{m'}$. 
We can further specify the frame by requiring that the flat space
gamma matrices $\Gamma_m$, $\Gamma_{m'}$ and $\Gamma_i$ are given by\footnote{%
  In any frame the gamma matrices are related to these by a similarity
  transformation $\Gamma_M \to S^{-1} \Gamma_M S = {\Lambda_M}^N \Gamma_N$. The
  (local) Lorentz transformation $\Lambda$ can be absorbed by a change of frame to ensure
  that the gamma matrices are in the desired basis.
}
\begin{align}
 \Gamma_m 
 =
 \gamma_m \otimes \rho_* \otimes \tau_*\,,
 \qquad
 \Gamma_{m'}
 =
 \mathds{1}_2 \otimes \rho_{m'} \otimes \tau_*\,,
 \qquad
 \Gamma_{i}
 =
 \mathds{1}_2 \otimes \mathds{1}_4 \otimes \tau_i\,,
\end{align}
with $\gamma_m$, $\rho_{m'}$ and $\tau_i$ respectively the gamma matrices for
$SO(1,2)$, $SO(4)_N$ and $SO(4)_S$, and $\rho_* \equiv \rho_1 \rho_2 \rho_3
\rho_4$, $\tau_* \equiv \tau_1 \tau_2 \tau_3 \tau_4$.
Note that $\rho_*^2 = \tau_*^2 = 1$.
With this basis the lagrangian becomes
\begin{equation}
\begin{aligned}
  &\bar{\Psi} \left( \gamma^p \rho_* \tau_* - \frac{1}{2} \gamma_{m n} \varepsilon^{mnp} \right)
  E_p^\mu
  \left (\partial_\mu \psi
 - \frac{1}{4} \gamma_{qr} \omega_\mu^{qr}
 - \frac{1}{2} \gamma_q \rho_* \rho_{r'} \omega_\mu^{qr'}
 - \frac{1}{4} \rho_{q'r'} \omega_\mu^{q'r'}
 \right) \Psi\\
 & \quad 
 + \frac{3}{R} \bar{\Psi} \left( 1 - \gamma_{012} \rho_* \tau_* \right) \tau_* \Psi\,.
\end{aligned}
\end{equation}
$\omega_\mu^{qr}$ and $\omega_\mu^{q'r'}$ are respectively the spin connections on
the world-volume and on the normal bundle. $\omega_{\mu}^{qr'}$ can be expressed as
\begin{align}
  \omega_\mu^{q r'}
  = E^{\sigma q} E_{\rho'}^{r'} \Gamma_{\mu\sigma}^{\rho'}
  = - E^{\sigma q} E_{\rho'}^{r'} \IIFundForm_{\mu\sigma}^{\rho'}\,,
\end{align}
where in the first equality we used the definition of the spin connection in
terms of a frame (analogous to~\eqref{eqn:nbspinconnection}), and in the second
we used that the second fundamental form~\eqref{SFF2} 
is also the Christoffel symbol for the metric $G$.

With some gamma matrix algebra and using that $\gamma_{012} = 1$, this can be rewritten as
\begin{equation}
\begin{aligned}
 &-\bar{\Psi}
 (1 - \rho_* \tau_*) \gamma^\mu \left(\partial_\mu - \frac{1}{4} \gamma_{mn}
 \omega_\mu^{mn}
 - \frac{1}{4} \rho_{m'n'} \omega_\mu^{m'n'} \right) \Psi
 +\frac{3}{2R} \bar{\Psi} (1 - \rho_* \tau_*) \tau_* \Psi\\
 & \quad - \frac{1}{2} \bar{\Psi} (1 - \rho_* \tau_*) \rho_* \rho_{m'} \Psi
 H^{m'}\,,
\end{aligned}
\end{equation}
and recall that the mean curvature vector $H$ vanishes because of the equations
of motion.
Finally, the M2-brane action has $\kappa$-symmetry, and we can fix a gauge by
imposing the condition
\begin{align}
  \left(1 + \frac{1}{6} E_{\mu}^m E_{\nu}^n E_\rho^p
  \Gamma_{mnp} \varepsilon^{\mu\nu\rho} \right) \Psi
  =
  \left( 1 + \rho_* \tau_* \right) \Psi = 0\,.
\end{align}
This condition is solved by decomposing $\Psi = \frac{1}{\sqrt{2}}(\psi_+ +
\psi_-)$ in terms of spinors $\psi_\pm$ with chirality
\begin{align}
  \tau_* \psi_\pm = -\rho _* \psi_{\pm} = \pm \psi_\pm.
\end{align}
We then obtain the gauge-fixed quadratic action
\begin{equation}
\begin{aligned}
 S_\text{fluc}^\text{(fer)}
 &=
 -T_\text{M2} \int_\Sigma \diff^3 \sigma\sqrt{-g}
 \left[ \bar{\psi}_+ \left( \slashed{D} + \frac{3}{2R} \right) \psi_+
 + \bar{\psi}_- \left( \slashed{D} - \frac{3}{2R} \right) \psi_- \right]\,,\\
 \slashed{D} &\equiv \gamma^\mu D_\mu
 = \gamma^\mu \left(\partial_\mu - \frac{1}{4} \gamma_{mn} \omega_\mu^{mn}
 - \frac{1}{4} \rho_{m'n'} \omega_\mu^{m'n'} \right)\,.
 \label{eqn:fermionic-det}
\end{aligned}
\end{equation}

We can read the relevant determinant for the analysis of section~\ref{sec:det} from
this action. After Wick rotation to euclidean signature, the path
integral over quadratic fluctuations gives the pfaffian
\begin{align}
  \Pf \left(\slashed{D} + \frac{3}{2R} \right) \Pf \left(\slashed{D} -
  \frac{3}{2R} \right)\,.
\end{align}
Using the identity
\begin{align}
  \Pf \left(\slashed{D} \pm \frac{3}{2R} \right)^2
  =
  \det \left(\slashed{D} \pm \frac{3}{2R} \right)
  =
  \det{}^{1/2} \left( - \slashed{D}^2 + \frac{9}{4R^2} \right)\,,
\end{align}
we can write
\begin{align}
  \Pf \left(\slashed{D} + \frac{3}{2R} \right) \Pf \left(\slashed{D} -
  \frac{3}{2R} \right)
  =
  \det{}^{1/2}\left( - \slashed{D}^2 + \frac{9}{4R^2} \right)\,,
  \label{eqn:fermionic-det2}
\end{align}
up to a possible overall sign ambiguity in the path integral.

We note that our gauge-fixed action~\eqref{eqn:fermionic-det} and the
determinant~\eqref{eqn:fermionic-det2} agree with existing results for fermionic
fluctuations derived for specific geometries: they reproduces that of the parallel
planes~\cite{Forste:1999yj} and the Dirac operator for $AdS_3$~\cite{Drukker:2020swu}.

\section{Convoluting the fermionic heat kernels}
\label{app:integral}

Here we evaluate the integral \eqref{eqn:spinorheatkernelrhointegral} 
(without the factor of $\tr_S(\mathds{1})$)
\begin{equation}
\begin{aligned}
 -4\pi
 & \int_0^t \diff t' \int_0^\infty \diff \rho \sinh^2{\rho}\,
 k_F(t-t';\rho) 
 \left[
 \frac{z_0^2}{3}\tr\bar\Pi
 \left( -\partial_\rho^2 - 2 \coth{\rho}\,\partial_\rho + \frac{1}{2} \tanh^2\frac{\rho}{2} -
 \frac{3}{2} \right) \right.\\
 & \qquad \left.
 + z_0^2 \bar{\RicciScalar}
 \left(
 \frac{1}{6} \frac{\cosh{\rho} \sinh{\rho} - \rho}{\sinh^2\rho} \partial_\rho
 + \frac{1}{4} - \frac{\rho + \sinh{\rho}}{24}
 \frac{\sinh(\rho/2)}{\cosh^3(\rho/2)}
 \right)
 \right]
 k_F(t';\rho) \,.
\end{aligned}
\end{equation}
The term multiplying $\bar\Pi$ is related to the heat kernel 
equation~\eqref{eqn:heatkernleqnspinor}. Then integrating by parts 
the first term on the second line we get
\begin{equation}
\begin{aligned}
 4\pi
 & \int_0^t \diff t' \int_0^\infty \diff \rho \sinh^2{\rho}\,
 k_F(t-t';\rho) 
 \\& \quad \times 
\left[
 \frac{z_0^2}{3}\tr\bar\Pi(\partial_t+\cM^{(0)})k_F(t';\rho) 
 -z_0^2 \bar{\RicciScalar}\frac{2-\rho\tanh(\rho/2)}{24\cosh^2(\rho/2)}
 k_F(t';\rho)
 \right].
\end{aligned}
\end{equation}
For the first term, we can apply the same
analysis as in the scalar case~\eqref{eqn:k2trPiint}, yielding
\begin{align}
 \frac{z_0^2}{3} \tr\bar{\Pi}\, t (\partial_t + \cM^{(0)}) k_F(t;0) \,.
 \label{eqn:k2ftrPiint}
\end{align}
While $k_F$ does not satisfy a general convolution, as it is not 
a heat kernel in 3d, this calculation still works because $U=1$ for 
the radial motion from 0 to $\rho$ and back to 0. It would be nice 
to try to repeat the calculation in this appendix using proper 
convolution, relying on the full 
fermionic heat kernel including angular dependence.

For the term proportional to $\bar\RicciScalar$, we can use the explicit
expression for the $AdS_3$ spinor heat kernel to evaluate the integral.
Plugging in $k_F$~\eqref{eqn:heatkernelspinorsol}, we get
\bal
  -\frac{z_0^2\bar\RicciScalar e^{-t\cM^{(0)}}}{384\pi^2} 
  \int_0^\infty \diff \rho\,
  \frac{2-\rho\tanh(\rho/2)}{\cosh^2(\rho/2)}
  \int_0^t \diff t'
  &\left(
\rho\left(\rho + t \tanh\frac{\rho}{2}\right)
\frac{e^{-t\rho^2/4t'(t-t')}}{(t'(t-t'))^{3/2}}
  \right.\\&\quad\left.
  +\tanh^2\frac{\rho}{2}
  \frac{e^{-t\rho^2/4t'(t-t')}}{(t'(t-t'))^{1/2}}
  \right).
  \label{eqn:appbintexplicit}
\eal
This can be evaluated using a small trick. We integrate 
the term with $(t'(t-t'))^{-1/2}$ by parts with respect to $\rho$ 
and note that the boundary term evaluates to zero, giving
\bal
  -\frac{z_0^2\bar\RicciScalar e^{-t\cM^{(0)}}}{1536\pi^2} 
  \int_0^\infty \diff \rho\,&
  \frac{\rho\left(2\rho(t+2)-(2\rho^2-3t)\sinh\rho+\rho(4-t)\cosh\rho\right)}
  {\cosh^4(\rho/2)}
  \int_0^t \diff t'
\frac{e^{-t\rho^2/4t'(t-t')}}{(t'(t-t'))^{3/2}}
\,.
\eal
Making a change of variables to $\tau=(t-2t')^2/4t'(t-t')$, 
the integral over $\tau$ is simply a gamma function and gives
\beq
\frac{z_0^2 \bar\RicciScalar e^{-t\cM^{(0)}}}{384 (\pi t)^{3/2}}
\int_0^\infty\diff\rho\,
\frac{(2\rho^2-3t)\sinh\rho-2 \rho  (t+2)+\rho  (t-4)
\cosh\rho}{\cosh^4(\rho/2)} e^{-(\rho^2/t)}\,.
\eeq
Finally the $\rho$ integral can be done straightforwardly and yields
\beq
- \frac{z_0^2\bar\RicciScalar}{96\pi^{3/2}} \frac{e^{-\cM^{(0)} t}}{\sqrt{t} }\,.
 \label{eqn:k2fricciint2}
\eeq

\bibliographystyle{utphys2}
\bibliography{ref}

\providecommand{\href}[2]{#2}\begingroup\raggedright\begin{thebibliography}{10}\setlength{\parskip}{1pt}\setlength{\itemsep}{0pt
  plus 0.3ex}

\bibitem{AJBray_1977}
A.~J. Bray and M.~A. Moore, ``Critical behaviour of semi-infinite systems,''
  \href{http://dx.doi.org/10.1088/0305-4470/10/11/021}{{\em J. Phys. A}
  {\bfseries 10} no.~11, (1977) 1927}.

\bibitem{ohno}
K.~Ohno and Y.~Okabe, ``{The $1/N$ expansion for the extraordinary transition
  of semi-infinite system},'' \href{http://dx.doi.org/10.1143/PTP.72.736}{{\em
  Prog. Th. Phys.} {\bfseries 72} no.~4, (1984) 736--745}.

\bibitem{Cardy:1984bb}
J.~L. Cardy, ``Conformal invariance and surface critical behavior,''
  \href{http://dx.doi.org/10.1016/0550-3213(84)90241-4}{{\em Nuclear Physics B}
  {\bfseries 240} no.~4, (1984) 514--532}.

\bibitem{Metlitski:2020cqy}
M.~A. Metlitski, ``{Boundary criticality of the $O(N)$ model in $d = 3$
  critically revisited},''
  \href{http://dx.doi.org/10.21468/SciPostPhys.12.4.131}{{\em SciPost Phys.}
  {\bfseries 12} no.~4, (2022) 131},
  \href{http://arxiv.org/abs/2009.05119}{{\ttfamily arXiv:2009.05119}}.

\bibitem{Rodriguez-Gomez:2022gbz}
D.~Rodriguez-Gomez, ``{A scaling limit for line and surface defects},''
  \href{http://dx.doi.org/10.1007/JHEP06(2022)071}{{\em JHEP} {\bfseries 06}
  (2022) 071}, \href{http://arxiv.org/abs/2202.03471}{{\ttfamily
  arXiv:2202.03471}}.

\bibitem{Shachar:2022fqk}
T.~Shachar, R.~Sinha, and M.~Smolkin, ``{RG flows on two-dimensional spherical
  defects},'' \href{http://arxiv.org/abs/2212.08081}{{\ttfamily
  arXiv:2212.08081}}.

\bibitem{Trepanier:2023tvb}
M.~Tr\'epanier, ``{Surface defects in the $O(N)$ model},''
  \href{http://dx.doi.org/10.1007/JHEP09(2023)074}{{\em JHEP} {\bfseries 09}
  (2023) 074}, \href{http://arxiv.org/abs/2305.10486}{{\ttfamily
  arXiv:2305.10486}}.

\bibitem{Giombi:2023dqs}
S.~Giombi and B.~Liu, ``{Notes on a surface defect in the $O(N)$ model},''
  \href{http://arxiv.org/abs/2305.11402}{{\ttfamily arXiv:2305.11402}}.

\bibitem{Raviv-Moshe:2023yvq}
A.~Raviv-Moshe and S.~Zhong, ``{Phases of surface defects in scalar field
  theories},'' \href{http://dx.doi.org/10.1007/JHEP08(2023)143}{{\em JHEP}
  {\bfseries 08} (2023) 143}, \href{http://arxiv.org/abs/2305.11370}{{\ttfamily
  arXiv:2305.11370}}.

\bibitem{Solodukhin:2008dh}
S.~N. Solodukhin, ``{Entanglement entropy, conformal invariance and extrinsic
  geometry},'' \href{http://dx.doi.org/10.1016/j.physletb.2008.05.071}{{\em
  Phys. Lett. B} {\bfseries 665} (2008) 305--309},
  \href{http://arxiv.org/abs/0802.3117}{{\ttfamily arXiv:0802.3117}}.

\bibitem{Bianchi:2015liz}
L.~Bianchi, M.~Meineri, R.~C. Myers, and M.~Smolkin, ``{R\'enyi entropy and
  conformal defects},'' \href{http://dx.doi.org/10.1007/JHEP07(2016)076}{{\em
  JHEP} {\bfseries 07} (2016) 076},
  \href{http://arxiv.org/abs/1511.06713}{{\ttfamily arXiv:1511.06713}}.

\bibitem{Gukov:2006jk}
S.~Gukov and E.~Witten, ``{Gauge theory, ramification, and the geometric
  Langlands program},'' \href{http://arxiv.org/abs/hep-th/0612073}{{\ttfamily
  hep-th/0612073}}.

\bibitem{Gukov:2008sn}
S.~Gukov and E.~Witten, ``{Rigid surface operators},''
  \href{http://dx.doi.org/10.4310/ATMP.2010.v14.n1.a3}{{\em Adv. Theor. Math.
  Phys.} {\bfseries 14} no.~1, (2010) 87--178},
  \href{http://arxiv.org/abs/0804.1561}{{\ttfamily arXiv:0804.1561}}.

\bibitem{Gomis:2014eya}
J.~Gomis and B.~Le~Floch, ``{M2-brane surface operators and gauge theory
  dualities in Toda},'' \href{http://dx.doi.org/10.1007/JHEP04(2016)183}{{\em
  JHEP} {\bfseries 04} (2016) 183},
  \href{http://arxiv.org/abs/1407.1852}{{\ttfamily arXiv:1407.1852}}.

\bibitem{Witten:1995zh}
E.~Witten, ``{Some comments on string dynamics},'' in {\em {STRINGS 95: Future
  Perspectives in String Theory}}, pp.~501--523.
\newblock 7, 1995.
\newblock \href{http://arxiv.org/abs/hep-th/9507121}{{\ttfamily
  hep-th/9507121}}.

\bibitem{Strominger:1995ac}
A.~Strominger, ``{Open $p$-branes},''
  \href{http://dx.doi.org/10.1016/0370-2693(96)00712-5}{{\em Phys. Lett. B}
  {\bfseries 383} (1996) 44--47},
  \href{http://arxiv.org/abs/hep-th/9512059}{{\ttfamily hep-th/9512059}}.

\bibitem{Ganor:1996nf}
O.~J. Ganor, ``{Six-dimensional tensionless strings in the large $N$ limit},''
  \href{http://dx.doi.org/10.1016/S0550-3213(96)00702-X}{{\em Nucl. Phys. B}
  {\bfseries 489} (1997) 95--121},
  \href{http://arxiv.org/abs/hep-th/9605201}{{\ttfamily hep-th/9605201}}.

\bibitem{Gaiotto:2009fs}
D.~Gaiotto, ``{Surface operators in $\mathcal{N} = 2$ 4d gauge theories},''
  \href{http://dx.doi.org/10.1007/JHEP11(2012)090}{{\em JHEP} {\bfseries 11}
  (2012) 090}, \href{http://arxiv.org/abs/0911.1316}{{\ttfamily
  arXiv:0911.1316}}.

\bibitem{Schwimmer:2008yh}
A.~Schwimmer and S.~Theisen, ``{Entanglement entropy, trace anomalies and
  holography},'' \href{http://dx.doi.org/10.1016/j.nuclphysb.2008.04.015}{{\em
  Nucl. Phys. B} {\bfseries 801} (2008) 1--24},
  \href{http://arxiv.org/abs/0802.1017}{{\ttfamily arXiv:0802.1017}}.

\bibitem{Wess:1971yu}
J.~Wess and B.~Zumino, ``{Consequences of anomalous Ward identities},''
  \href{http://dx.doi.org/10.1016/0370-2693(71)90582-X}{{\em Phys. Lett. B}
  {\bfseries 37} (1971) 95--97}.

\bibitem{Deser:1993yx}
S.~Deser and A.~Schwimmer, ``{Geometric classification of conformal anomalies
  in arbitrary dimensions},''
  \href{http://dx.doi.org/10.1016/0370-2693(93)90934-A}{{\em Phys. Lett. B}
  {\bfseries 309} (1993) 279--284},
  \href{http://arxiv.org/abs/hep-th/9302047}{{\ttfamily hep-th/9302047}}.

\bibitem{Boulanger:2007st}
N.~Boulanger, ``{General solutions of the Wess-Zumino consistency condition for
  the Weyl anomalies},''
  \href{http://dx.doi.org/10.1088/1126-6708/2007/07/069}{{\em JHEP} {\bfseries
  07} (2007) 069}, \href{http://arxiv.org/abs/0704.2472}{{\ttfamily
  arXiv:0704.2472}}.

\bibitem{Drukker:2020dcz}
N.~Drukker, M.~Probst, and M.~Tr\'epanier, ``{Surface operators in the 6d
  $\mathcal{N} = (2,0)$ theory},''
  \href{http://dx.doi.org/10.1088/1751-8121/aba1b7}{{\em J. Phys. A} {\bfseries
  53} no.~36, (2020) 365401}, \href{http://arxiv.org/abs/2003.12372}{{\ttfamily
  arXiv:2003.12372}}.

\bibitem{graham:1999pm}
C.~R. Graham and E.~Witten, ``{Conformal anomaly of submanifold observables in
  $AdS$/CFT correspondence},''
  \href{http://dx.doi.org/10.1016/S0550-3213(99)00055-3}{{\em Nucl. Phys.}
  {\bfseries B546} (1999) 52--64},
  \href{http://arxiv.org/abs/hep-th/9901021}{{\ttfamily hep-th/9901021}}.

\bibitem{Maldacena:1998im}
J.~M. Maldacena, ``{Wilson loops in large $N$ field theories},''
  \href{http://dx.doi.org/10.1103/PhysRevLett.80.4859}{{\em Phys. Rev. Lett.}
  {\bfseries 80} (1998) 4859--4862},
  \href{http://arxiv.org/abs/hep-th/9803002}{{\ttfamily hep-th/9803002}}.

\bibitem{Berenstein:1998ij}
D.~E. Berenstein, R.~Corrado, W.~Fischler, and J.~M. Maldacena, ``{The operator
  product expansion for Wilson loops and surfaces in the large $N$ limit},''
  \href{http://dx.doi.org/10.1103/PhysRevD.59.105023}{{\em Phys. Rev. D}
  {\bfseries 59} (1999) 105023},
  \href{http://arxiv.org/abs/hep-th/9809188}{{\ttfamily hep-th/9809188}}.

\bibitem{mezei:2018url}
M.~Mezei, S.~S. Pufu, and Y.~Wang, ``{Chern-Simons theory from M5-branes and
  calibrated M2-branes},''
  \href{http://dx.doi.org/10.1007/JHEP08(2019)165}{{\em JHEP} {\bfseries 08}
  (2019) 165}, \href{http://arxiv.org/abs/1812.07572}{{\ttfamily
  arXiv:1812.07572}}.

\bibitem{Drukker:2021vyx}
N.~Drukker and M.~Tr\'epanier, ``{M2-doughnuts},''
  \href{http://dx.doi.org/10.1007/JHEP02(2022)071}{{\em JHEP} {\bfseries 02}
  (2022) 071}, \href{http://arxiv.org/abs/2111.09385}{{\ttfamily
  arXiv:2111.09385}}.

\bibitem{Drukker:2022beq}
N.~Drukker and M.~Tr\'epanier, ``{Ironing out the crease},''
  \href{http://dx.doi.org/10.1007/JHEP08(2022)193}{{\em JHEP} {\bfseries 08}
  (2022) 193}, \href{http://arxiv.org/abs/2204.12627}{{\ttfamily
  arXiv:2204.12627}}.

\bibitem{Drukker:2022kuz}
N.~Drukker and M.~Tr\'epanier, ``{BPS surface operators and calibrations},''
  \href{http://dx.doi.org/10.1088/1751-8121/acc771}{{\em J. Phys. A} {\bfseries
  56} no.~17, (2023) 175403}, \href{http://arxiv.org/abs/2210.07251}{{\ttfamily
  arXiv:2210.07251}}.

\bibitem{Jensen:2018rxu}
K.~Jensen, A.~O'Bannon, B.~Robinson, and R.~Rodgers, ``{From the Weyl anomaly
  to entropy of two-dimensional boundaries and defects},''
  \href{http://dx.doi.org/10.1103/PhysRevLett.122.241602}{{\em Phys. Rev.
  Lett.} {\bfseries 122} no.~24, (2019) 241602},
  \href{http://arxiv.org/abs/1812.08745}{{\ttfamily arXiv:1812.08745}}.

\bibitem{Gentle:2015jma}
S.~A. Gentle, M.~Gutperle, and C.~Marasinou, ``{Entanglement entropy of Wilson
  surfaces from bubbling geometries in M-theory},''
  \href{http://dx.doi.org/10.1007/JHEP08(2015)019}{{\em JHEP} {\bfseries 08}
  (2015) 019}, \href{http://arxiv.org/abs/1506.00052}{{\ttfamily
  arXiv:1506.00052}}.

\bibitem{Rodgers:2018mvq}
R.~Rodgers, ``{Holographic entanglement entropy from probe M-theory branes},''
  \href{http://dx.doi.org/10.1007/JHEP03(2019)092}{{\em JHEP} {\bfseries 03}
  (2019) 092}, \href{http://arxiv.org/abs/1811.12375}{{\ttfamily
  arXiv:1811.12375}}.

\bibitem{Estes:2018tnu}
J.~Estes, D.~Krym, A.~O'Bannon, B.~Robinson, and R.~Rodgers, ``{Wilson surface
  central charge from holographic entanglement entropy},''
  \href{http://dx.doi.org/10.1007/JHEP05(2019)032}{{\em JHEP} {\bfseries 05}
  (2019) 032},
\href{http://arxiv.org/abs/1812.00923}{{\ttfamily arXiv:1812.00923}}.

\bibitem{Wang:2020xkc}
Y.~Wang, ``{Surface defect, anomalies and $b$-extremization},''
  \href{http://dx.doi.org/10.1007/JHEP11(2021)122}{{\em JHEP} {\bfseries 11}
  (2021) 122}, \href{http://arxiv.org/abs/2012.06574}{{\ttfamily
  arXiv:2012.06574}}.

\bibitem{Bullimore:2014upa}
M.~Bullimore and H.-C. Kim, ``{The superconformal index of the (2,0) theory
  with defects},'' \href{http://dx.doi.org/10.1007/JHEP05(2015)048}{{\em JHEP}
  {\bfseries 05} (2015) 048}, \href{http://arxiv.org/abs/1412.3872}{{\ttfamily
  arXiv:1412.3872}}.

\bibitem{Chalabi:2020iie}
A.~Chalabi, A.~O'Bannon, B.~Robinson, and J.~Sisti, ``{Central charges of 2d
  superconformal defects},''
  \href{http://dx.doi.org/10.1007/JHEP05(2020)095}{{\em JHEP} {\bfseries 05}
  (2020) 095}, \href{http://arxiv.org/abs/2003.02857}{{\ttfamily
  arXiv:2003.02857}}.

\bibitem{Meneghelli:2022gps}
C.~Meneghelli and M.~Tr\'epanier, ``{Bootstrapping string dynamics in the 6d
  $\mathcal{N} = (2, 0)$ theories},''
  \href{http://dx.doi.org/10.1007/JHEP07(2023)165}{{\em JHEP} {\bfseries 07}
  (2023) 165}, \href{http://arxiv.org/abs/2212.05020}{{\ttfamily
  arXiv:2212.05020}}.

\bibitem{Bianchi:2019sxz}
L.~Bianchi and M.~Lemos, ``{Superconformal surfaces in four dimensions},''
  \href{http://dx.doi.org/10.1007/JHEP06(2020)056}{{\em JHEP} {\bfseries 06}
  (2020) 056}, \href{http://arxiv.org/abs/1911.05082}{{\ttfamily
  arXiv:1911.05082}}.

\bibitem{Drukker:2020atp}
N.~Drukker, M.~Probst, and M.~Tr\'epanier, ``{Defect CFT techniques in the 6d
  $\mathcal{N} = (2,0)$ theory},''
  \href{http://dx.doi.org/10.1007/JHEP03(2021)261}{{\em JHEP} {\bfseries 03}
  (2021) 261}, \href{http://arxiv.org/abs/2009.10732}{{\ttfamily
  arXiv:2009.10732}}.

\bibitem{Corrado:1999pi}
R.~Corrado, B.~Florea, and R.~McNees, ``{Correlation functions of operators and
  Wilson surfaces in the $d=6$, (0,2) theory in the large $N$ limit},''
  \href{http://dx.doi.org/10.1103/PhysRevD.60.085011}{{\em Phys. Rev. D}
  {\bfseries 60} (1999) 085011},
  \href{http://arxiv.org/abs/hep-th/9902153}{{\ttfamily hep-th/9902153}}.

\bibitem{Drukker:2020swu}
N.~Drukker, S.~Giombi, A.~A. Tseytlin, and X.~Zhou, ``{Defect CFT in the 6d
  $(2,0)$ theory from M2 brane dynamics in $AdS_7 \times S^4$},''
  \href{http://dx.doi.org/10.1007/JHEP07(2020)101}{{\em JHEP} {\bfseries 07}
  (2020) 101}, \href{http://arxiv.org/abs/2004.04562}{{\ttfamily
  arXiv:2004.04562}}.

\bibitem{Vassilevich:2003xt}
D.~V. Vassilevich, ``{Heat kernel expansion: user's manual},''
  \href{http://dx.doi.org/10.1016/j.physrep.2003.09.002}{{\em Phys. Rept.}
  {\bfseries 388} (2003) 279--360},
  \href{http://arxiv.org/abs/hep-th/0306138}{{\ttfamily hep-th/0306138}}.

\bibitem{Camporesi:1990wm}
R.~Camporesi, ``{Harmonic analysis and propagators on homogeneous spaces},''
  \href{http://dx.doi.org/10.1016/0370-1573(90)90120-Q}{{\em Phys. Rept.}
  {\bfseries 196} (1990) 1--134}.

\bibitem{Giombi:2008vd}
S.~Giombi, A.~Maloney, and X.~Yin, ``{One-loop partition functions of 3D
  gravity},'' \href{http://dx.doi.org/10.1088/1126-6708/2008/08/007}{{\em JHEP}
  {\bfseries 08} (2008) 007}, \href{http://arxiv.org/abs/0804.1773}{{\ttfamily
  arXiv:0804.1773}}.

\bibitem{Giombi:2013fka}
S.~Giombi and I.~R. Klebanov, ``{One loop tests of higher spin AdS/CFT},''
  \href{http://dx.doi.org/10.1007/JHEP12(2013)068}{{\em JHEP} {\bfseries 12}
  (2013) 068}, \href{http://arxiv.org/abs/1308.2337}{{\ttfamily
  arXiv:1308.2337}}.

\bibitem{Forini:2017whz}
V.~Forini, A.~A. Tseytlin, and E.~Vescovi, ``{Perturbative computation of
  string one-loop corrections to Wilson loop minimal surfaces in $AdS_5 \times
  S^5$},'' \href{http://dx.doi.org/10.1007/JHEP03(2017)003}{{\em JHEP}
  {\bfseries 03} (2017) 003}, \href{http://arxiv.org/abs/1702.02164}{{\ttfamily
  arXiv:1702.02164}}.

\bibitem{fefferman1985conformal}
C.~Fefferman, ``{Conformal invariants},'' {\em ``Elie Cartan et les
  mathématiques d'aujourd'hui,'' Astérisque, hors série} (1985) 95--116.

\bibitem{Bergshoeff:1987cm}
E.~Bergshoeff, E.~Sezgin, and P.~K. Townsend, ``{Supermembranes and
  eleven-dimensional supergravity},''
\href{http://dx.doi.org/10.1016/0370-2693(87)91272-X}{{\em Phys. Lett.}
  {\bfseries B189} (1987) 75--78}.

\bibitem{Drukker:1999zq}
N.~Drukker, D.~J. Gross, and H.~Ooguri, ``{Wilson loops and minimal
  surfaces},'' \href{http://dx.doi.org/10.1103/PhysRevD.60.125006}{{\em Phys.
  Rev. D} {\bfseries 60} (1999) 125006},
  \href{http://arxiv.org/abs/hep-th/9904191}{{\ttfamily hep-th/9904191}}.

\bibitem{Mori:2014tca}
H.~Mori and S.~Yamaguchi, ``{M5-branes and Wilson surfaces in
  $AdS_{7}$/CFT$_{6}$ correspondence},''
  \href{http://dx.doi.org/10.1103/PhysRevD.90.026005}{{\em Phys. Rev. D}
  {\bfseries 90} no.~2, (2014) 026005},
  \href{http://arxiv.org/abs/1404.0930}{{\ttfamily 1404.0930}}.

\bibitem{lawson}
H.~B. Lawson and M.-L. Michelsohn, {\em Spin Geometry and the Dirac Operators},
  pp.~77--165.
\newblock Princeton University Press, 1989.
\newblock \url{http://www.jstor.org/stable/j.ctt1bpmb28.7}.

\bibitem{Camporesi:1992tm}
R.~Camporesi, ``{The spinor heat kernel in maximally symmetric spaces},''
  \href{http://dx.doi.org/10.1007/BF02100862}{{\em Commun. Math. Phys.}
  {\bfseries 148} (1992) 283--308}.

\bibitem{Camporesi:1995fb}
R.~Camporesi and A.~Higuchi, ``{On the eigen functions of the Dirac operator on
  spheres and real hyperbolic spaces},''
  \href{http://dx.doi.org/10.1016/0393-0440(95)00042-9}{{\em J. Geom. Phys.}
  {\bfseries 20} (1996) 1--18},
  \href{http://arxiv.org/abs/gr-qc/9505009}{{\ttfamily gr-qc/9505009}}.

\bibitem{David:2009xg}
J.~R. David, M.~R. Gaberdiel, and R.~Gopakumar, ``{The heat kernel on $AdS_3$
  and its applications},''
  \href{http://dx.doi.org/10.1007/JHEP04(2010)125}{{\em JHEP} {\bfseries 04}
  (2010) 125}, \href{http://arxiv.org/abs/0911.5085}{{\ttfamily
  arXiv:0911.5085}}.

\bibitem{Bergamin:2015vxa}
R.~Bergamin and A.~A. Tseytlin, ``{Heat kernels on cone of $AdS_2$ and
  $k$-wound circular Wilson loop in $AdS_5 \times S^5$ superstring},''
  \href{http://dx.doi.org/10.1088/1751-8113/49/14/14LT01}{{\em J. Phys. A}
  {\bfseries 49} no.~14, (2016) 14LT01},
  \href{http://arxiv.org/abs/1510.06894}{{\ttfamily arXiv:1510.06894}}.

\bibitem{Forste:1999yj}
{F\"orste, Stefan}, ``{Membrany corrections to the string anti-string potential
  in M5-brane theory},''
  \href{http://dx.doi.org/10.1088/1126-6708/1999/05/002}{{\em JHEP} {\bfseries
  05} (1999) 002}, \href{http://arxiv.org/abs/hep-th/9902068}{{\ttfamily
  hep-th/9902068}}.

\bibitem{Giombi:2023vzu}
S.~Giombi and A.~A. Tseytlin, ``{Wilson loops at large $N$ and the quantum
  M2-brane},'' \href{http://dx.doi.org/10.1103/PhysRevLett.130.201601}{{\em
  Phys. Rev. Lett.} {\bfseries 130} no.~20, (2023) 201601},
  \href{http://arxiv.org/abs/2303.15207}{{\ttfamily arXiv:2303.15207}}.

\bibitem{Beccaria:2023ujc}
M.~Beccaria, S.~Giombi, and A.~A. Tseytlin, ``{Instanton contributions to the
  ABJM free energy from quantum M2 branes},''
  \href{http://dx.doi.org/10.1007/JHEP10(2023)029}{{\em JHEP} {\bfseries 10}
  (2023) 029}, \href{http://arxiv.org/abs/2307.14112}{{\ttfamily
  arXiv:2307.14112}}.

\bibitem{Seibold:2023zkz}
F.~K. Seibold and A.~A. Tseytlin, ``{S-matrix on effective string and
  compactified membrane},''
  \href{http://dx.doi.org/10.1088/1751-8121/ad05f0}{{\em J. Phys. A} {\bfseries
  56} no.~48, (2023) 485401}, \href{http://arxiv.org/abs/2308.12189}{{\ttfamily
  arXiv:2308.12189}}.

\bibitem{Beccaria:2023sph}
M.~Beccaria, S.~Giombi, and A.~A. Tseytlin, ``{(2,0) theory on $S^5 \times S^1$
  and quantum M2 branes},'' \href{http://arxiv.org/abs/2309.10786}{{\ttfamily
  arXiv:2309.10786}}.

\bibitem{toappear}
N.~Drukker and O.~Shahpo. To appear.

\bibitem{Drukker:2020bes}
N.~Drukker and M.~Tr\'epanier, ``{Observations on BPS observables in 6d},''
  \href{http://dx.doi.org/10.1088/1751-8121/abf38d}{{\em J. Phys. A} {\bfseries
  54} no.~20, (2021) 20}, \href{http://arxiv.org/abs/2012.11087}{{\ttfamily
  arXiv:2012.11087}}.

\bibitem{Polchinski:2011im}
J.~Polchinski and J.~Sully, ``{Wilson loop renormalization group flows},''
  \href{http://dx.doi.org/10.1007/JHEP10(2011)059}{{\em JHEP} {\bfseries 10}
  (2011) 059}, \href{http://arxiv.org/abs/1104.5077}{{\ttfamily
  arXiv:1104.5077}}.

\bibitem{Beccaria:2017rbe}
M.~Beccaria, S.~Giombi, and A.~Tseytlin, ``{Non-supersymmetric Wilson loop in $
  \mathcal{N} = 4$ SYM and defect 1d CFT},''
  \href{http://dx.doi.org/10.1007/JHEP03(2018)131}{{\em JHEP} {\bfseries 03}
  (2018) 131}, \href{http://arxiv.org/abs/1712.06874}{{\ttfamily
  arXiv:1712.06874}}.

\bibitem{Chen:2007ir}
B.~Chen, W.~He, J.-B. Wu, and L.~Zhang, ``{M5-branes and Wilson surfaces},''
  \href{http://dx.doi.org/10.1088/1126-6708/2007/08/067}{{\em JHEP} {\bfseries
  08} (2007) 067}, \href{http://arxiv.org/abs/0707.3978}{{\ttfamily
  arXiv:0707.3978}}.

\bibitem{Lewkowycz:2014jia}
A.~Lewkowycz and E.~Perlmutter, ``{Universality in the geometric dependence of
  Rényi entropy},'' \href{http://dx.doi.org/10.1007/JHEP01(2015)080}{{\em
  JHEP} {\bfseries 01} (2015) 080},
  \href{http://arxiv.org/abs/1407.8171}{{\ttfamily arXiv:1407.8171}}.

\bibitem{Drukker:2008wr}
N.~Drukker, J.~Gomis, and S.~Matsuura, ``{Probing ${\cal N}=4$ SYM with surface
  operators},'' \href{http://dx.doi.org/10.1088/1126-6708/2008/10/048}{{\em
  JHEP} {\bfseries 10} (2008) 048},
  \href{http://arxiv.org/abs/0805.4199}{{\ttfamily arXiv:0805.4199}}.

\bibitem{Chalabi:2021jud}
A.~Chalabi, C.~P. Herzog, A.~O'Bannon, B.~Robinson, and J.~Sisti, ``{Weyl
  anomalies of four dimensional conformal boundaries and defects},''
  \href{http://dx.doi.org/10.1007/JHEP02(2022)166}{{\em JHEP} {\bfseries 02}
  (2022) 166}, \href{http://arxiv.org/abs/2111.14713}{{\ttfamily
  arXiv:2111.14713}}.

\bibitem{Capuozzo:2023fll}
P.~Capuozzo, J.~Estes, B.~Robinson, and B.~Suzzoni, ``{Holographic Weyl
  anomalies for 4d defects in 6d SCFTs},''
  \href{http://arxiv.org/abs/2310.17447}{{\ttfamily arXiv:2310.17447}}.

\bibitem{Drukker:2000ep}
N.~Drukker, D.~J. Gross, and A.~A. Tseytlin, ``{Green-Schwarz string in
  $AdS_5\times S^5$: semiclassical partition function},''
  \href{http://dx.doi.org/10.1088/1126-6708/2000/04/021}{{\em JHEP} {\bfseries
  04} (2000) 021}, \href{http://arxiv.org/abs/hep-th/0001204}{{\ttfamily
  hep-th/0001204}}.

\bibitem{Forini:2015mca}
V.~Forini, V.~G.~M. Puletti, L.~Griguolo, D.~Seminara, and E.~Vescovi,
  ``{Remarks on the geometrical properties of semiclassically quantized
  strings},'' \href{http://dx.doi.org/10.1088/1751-8113/48/47/475401}{{\em J.
  Phys. A} {\bfseries 48} no.~47, (2015) 475401},
  \href{http://arxiv.org/abs/1507.01883}{{\ttfamily arXiv:1507.01883}}.

\bibitem{deLeonArdon:2020crs}
R.~de~Le\'on~Ard\'on, ``{Semiclassical $p$-branes in hyperbolic space},''
  \href{http://dx.doi.org/10.1088/1361-6382}{{\em Class. Quant. Grav.}
  {\bfseries 37} no.~23, (2020) 237001},
  \href{http://arxiv.org/abs/2007.03591}{{\ttfamily arXiv:2007.03591}}.

\bibitem{Goon:2020myi}
G.~Goon, S.~Melville, and J.~Noller, ``{Quantum corrections to generic branes:
  DBI, NLSM, and more},'' \href{http://dx.doi.org/10.1007/JHEP01(2021)159}{{\em
  JHEP} {\bfseries 01} (2021) 159},
  \href{http://arxiv.org/abs/2010.05913}{{\ttfamily arXiv:2010.05913}}.

\bibitem{deWit:1998tk}
B.~de~Wit, K.~Peeters, and J.~Plefka, ``{Superspace geometry for supermembrane
  backgrounds},'' \href{http://dx.doi.org/10.1016/S0550-3213(98)00445-3}{{\em
  Nucl. Phys. B} {\bfseries 532} (1998) 99--123},
  \href{http://arxiv.org/abs/hep-th/9803209}{{\ttfamily hep-th/9803209}}.

\bibitem{deWit:1998yu}
B.~de~Wit, K.~Peeters, J.~Plefka, and A.~Sevrin, ``{The M-theory two-brane in
  $AdS_4 \times S^7$ and $AdS_7 \times S^4$},''
  \href{http://dx.doi.org/10.1016/S0370-2693(98)01340-9}{{\em Phys. Lett. B}
  {\bfseries 443} (1998) 153--158},
  \href{http://arxiv.org/abs/hep-th/9808052}{{\ttfamily hep-th/9808052}}.

\bibitem{Sakaguchi:2010dg}
M.~Sakaguchi, H.~Shin, and K.~Yoshida, ``{Semiclassical analysis of M2-brane in
  $AdS_4 \times S^7 / Z_k$},''
  \href{http://dx.doi.org/10.1007/JHEP12(2010)012}{{\em JHEP} {\bfseries 12}
  (2010) 012}, \href{http://arxiv.org/abs/1007.3354}{{\ttfamily
  arXiv:1007.3354}}.

\end{thebibliography}\endgroup

\end{document}